\documentclass[12pt,a4paper]{article}

\usepackage{amsthm,amsmath,amsfonts,amssymb}
\usepackage{a4,times,latexsym}
\usepackage{eurosym} 

\usepackage{subfig} 
\usepackage{graphicx} 
\usepackage[countmax]{subfloat} 

\usepackage{nohyperref}
\usepackage{xcolor}
\hypersetup{
    colorlinks,
    linkcolor={red!50!black},
    citecolor={blue!50!black},
    urlcolor={blue!80!black}
}

\usepackage{lscape} 
\usepackage{epstopdf} 
\usepackage{listings} 
\usepackage{setspace} 
\usepackage{url}      
\usepackage{scrpage2} 
\usepackage{makeidx}  
\usepackage{rotating,here} 

\usepackage{lineno} 

\usepackage{tabularx} 
\usepackage{booktabs} 
\newcolumntype{Y}{>{\raggedleft\arraybackslash}X} 

\usepackage{color}
\usepackage{psfrag}
\usepackage{lmodern}
\usepackage[USenglish]{babel}
\usepackage[T1]{fontenc}
\usepackage[latin1]{inputenc}
\usepackage[round,comma]{natbib}
\usepackage{geometry}
\theoremstyle{plain}

\theoremstyle{definition}

\usepackage{tikz,color}
\usepackage{pgflibraryshapes}
\usetikzlibrary{trees,arrows, decorations.markings, positioning}
\definecolor{grau}{rgb}{0.9,0.9,0.9}
\tikzstyle{vecArrow} = [line width=2mm, draw=black, -triangle 60,
                        postaction = {draw, line width=5mm, shorten >=6mm, -},
                        preaction = {decorate}]
\tikzstyle{innerWhite} = [line width=1.9mm, draw=grau, -triangle 60,
                        postaction = {draw, line width=4.5mm, shorten >=5.5mm, -}]

\begin{document}

\renewcommand{\thefootnote}{\arabic{footnote}}

\begin{center}
\large{\bf Using the lasso method for space-time short-term wind speed predictions\\}
\vspace{10mm} \normalsize \today
\small

Daniel Ambach$^1$ and Carsten Croonenbroeck$^2$ 

\end{center}

\thispagestyle{empty} \small

\normalsize

\begin{abstract}
\noindent Accurate wind power forecasts depend on reliable wind speed
forecasts. Numerical Weather Predictions (NWPs) utilize huge amounts
of computing time, but still have rather low spatial and temporal
resolution. However, stochastic wind speed forecasts perform well in
rather high temporal resolution settings. They consume comparably
little computing resources and return reliable forecasts, if
forecasting horizons are not too long. In the recent literature,
spatial interdependence is increasingly taken into consideration. In
this paper we propose a new and quite flexible multivariate model
that accounts for neighbouring weather stations' information and as
such, exploits spatial data at a high resolution. The model is
applied to forecasting horizons of up to one day and is capable of
handling a high resolution temporal structure. We use a periodic
vector autoregressive model with seasonal lags to account for the
interaction of the explanatory variables. Periodicity is considered
and is modelled by cubic B-splines. Due to the model's flexibility, the
number of explanatory variables becomes huge. Therefore, we utilize time-saving shrinkage methods like lasso and elastic net for estimation. Particularly, a
relatively newly developed iteratively re-weighted lasso and elastic net is applied
that also incorporates heteroscedasticity. We compare our model to
several benchmarks. The out-of-sample forecasting results show that the
exploitation of spatial information increases the forecasting accuracy tremendously, in comparison to
models in use so far.
\end{abstract}

\begin{small}{\bf Keywords:} Wind speed; Forecasting; Periodic vector autoregressive model; Periodic B-splines; Iteratively re-weighted lasso method
\end{small}

\bigskip

\begin{bfseries}
Addresses:
\end{bfseries}

\begin{small}
$^{1}$ \textbf{Corresponding Author:} Daniel Ambach, European University Via\-dri\-na, Chair of Quantitative Methods and Statistics,
Post Box 1786, 15207 Frankfurt (Oder), Germany, Tel. +49 (0)335 5534 2983, Fax +49 (0)335 5534 2233, ambach@europa-uni.de.\vspace{0.1cm}

$^2$ Carsten Croonenbroeck,
European University Via\-dri\-na, Chair of Economics and Economic Theory (Macroeconomics),
Post Box 1786, 15207 Frankfurt (Oder), Germany, Tel. +49 (0)335 5534 2701, Fax +49 (0)335 5534 72701, croonenbroeck@europa-uni.de.\vspace{0.5cm}
\end{small}

\pagestyle{plain}
\setcounter{page}{1}
\onehalfspacing

\section{Introduction}\label{section:intro}

The progressing energy turnaround in Europe has one designated goal:
Reducing the dependence on fossil energy and thus, increasing the
fraction of energy production that is obtained from renewable
sources. Other renewables like solar power or geothermal power
aside, wind power is the most successful one, in terms of installed
capacity as well as in the aggregated amount of power production. Contrary to fossil
power however, wind power production is erratic and
non-deterministic. Thus, accurate predictions are crucial for
efficient market clearing as well as network dispatching, as
\cite{Soman2010} and \cite{Croonenbroeck2014} point out. Short- to
medium-term wind power forecasting (up to one day) is a wide field
of research. These predictions are usually carried out by stochastic
approaches, as discussed by \cite{Wu2007}.\\
Wind power prediction depends basically on wind speed predictions. However, wind power forecasting is a task on its own, i.e. power production depends on the turbine type, among other things, as \cite{Burton2011} indicate. Wind speed prediction models have been under constant development and improvement during the recent decade, at least. Early work is done by \cite{Haslett1989}. They propose a long-memory autoregressive moving average model for wind speed. \cite{Ewing2006} investigate heteroscedasticity of high frequency data (15-minute observation frequency). Thus, \cite{Koopman2007} and \cite{taylor2009wind} apply an ARFIMA-GARCH model (autoregressive fractionally integrated moving average with generalized autoregressive conditional heteroscedasticity) with a seasonal component in the explanatory variables. \\
Instead of univariate processes, more recently, researchers exploit spatial information of the data, as weather stations are available at close proximity to each other in many cases. \cite{gneiting2006calibrated} develop a spatial regime-switching model. Furthermore, \cite{zhu2014space} provide an extended regime-switching approach. \cite{vsaltyte2011} model the spatial dependence of the daily wind speed by a Gaussian random field. \cite{Aguera-Perez2013} and \cite{Santos-Alamillos2014} investigate the spatial structure of the data as well.\\
In this paper we come up with a very general and flexible model that uses spatial wind speed, wind direction, temperature and air pressure data. The interaction of the data is modelled by a periodic vector autoregressive model (SVAR) with explanatory variables (referred to as ``X''), which are, in our case, just several periodic regressors. Contrary to classical Fourier modeling for the periodicity, we use more flexible B-splines. The heteroscedastic variance is modelled by a threshold autoregressive conditional heteroscedasticity (TARCH) model, also with explanatory variables (which are again, in our case, periodic regressors). In the end, our model is an SVARX-TARCHX model with adaptive lag selection. 
Due to the huge parameter space, classical maximum likelihood (ML) estimation would consume a lot of computing time. Instead, we use the popular shrinkage methodology, applied by means of the elastic net \citep[see][]{zou2005regularization} and lasso (least absolute selection and shrinkage operator) method \citep[see][]{tibshirani1996}. In the context of autoregressive and time series models, \cite{ren2010}, \cite{rajapakse2011stability} and \cite{yoon2013penalized} apply lasso type and elastic net methods. \cite{evans2014} use the lasso and several other empirical models to enhance the forecasting performance of a wind farm. The lasso and the elastic net are efficient ways to perform model selection and model fitting in one step. These algorithms operate quite fast and do not require any distributional assumption. Recently, \cite{ziel2014} proposed an iteratively re-weighted lasso method which incorporates heteroscedasticity within the variance part.\\
We apply our model to a data set that consists of seven weather stations in Germany. The stations record weather data at a frequency of ten minutes. Per station, we have three full sample years, i.e. roughly 158,000 observations. With this data set, we calculate forecasts from our model and from several benchmark models, i.e. the persistence model, an AR, a VAR model and an ARFIMA-APARCH model (ARFIMA with asymmetric power generalized autoregressive conditional heteroscedasticity), as applied by \cite{Ambach2014}. The novel multivariate approach is calculated by two different estimation procedures, the lasso method and the elastic net. We compare all models according to their forecasting accuracy and give a proposal of the  best modeling approach.\\
The article is structured as follows. In Section \ref{section:empirics}, we briefly describe the data set and some general properties of the data set. Furthermore, the univariate and the benchmark models are described. The novel multivariate approach is introduced. Section \ref{section:insample} describes the in-sample results of our new modeling approach. The out-of sample results are provided in Section \ref{section:oosample}. Finally, Section \ref{section:conclusion} concludes.

\section{Wind speed data and model description}\label{section:empirics}

\begin{figure}[h]
  \includegraphics[width=.7\textwidth,trim= 2.05cm 2.55cm 1cm 2cm,clip=true]{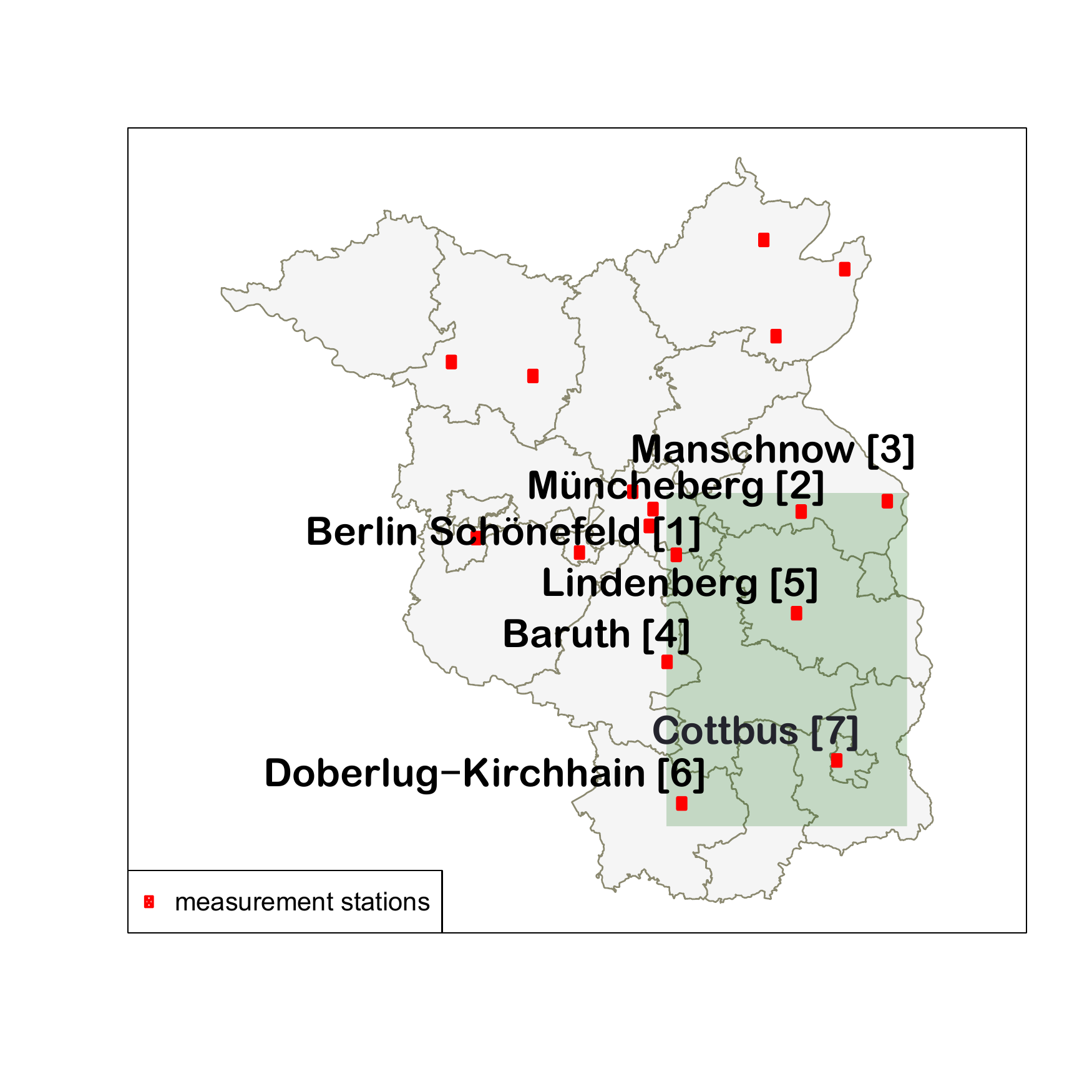}
  \caption{Meteorological measurement stations in Berlin and Brandenburg which provide 10min data.}\label{figure:brand}
\end{figure}

\begin{figure}[h]
  \includegraphics[width=1\textwidth,trim= .1cm .55cm .5cm .1cm,clip=true]{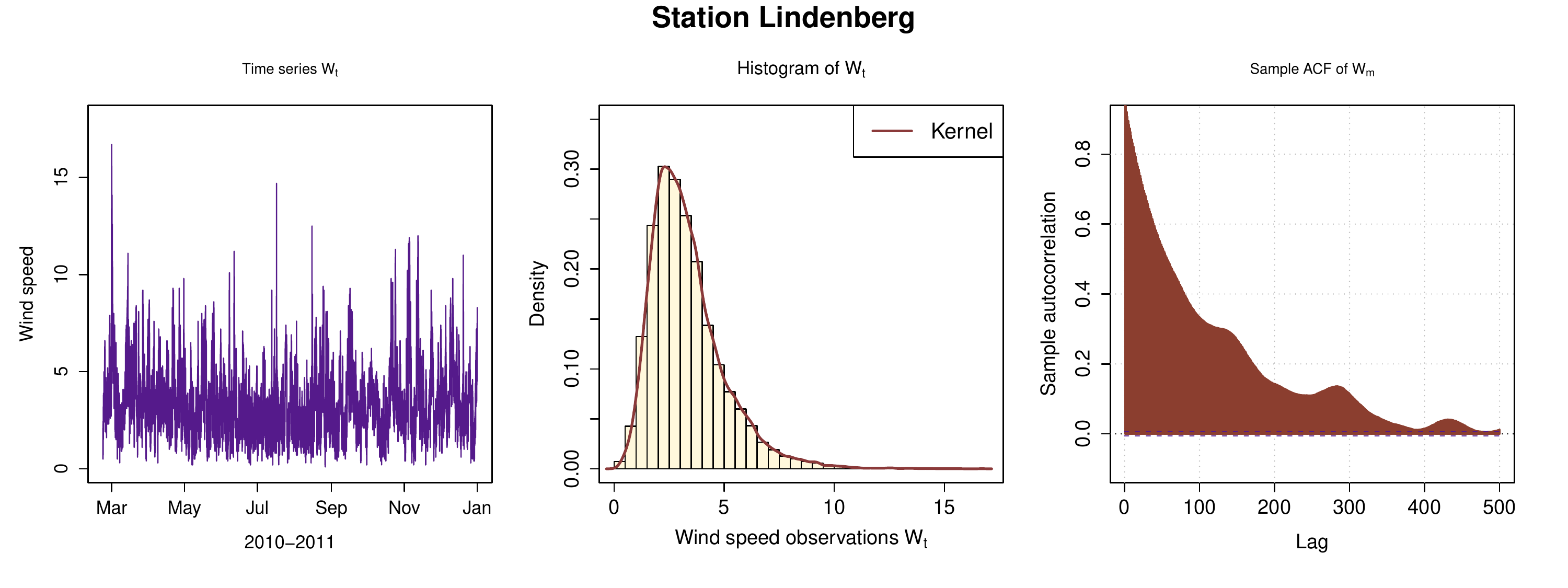}
  \includegraphics[width=1\textwidth,trim= .1cm .55cm .5cm .1cm,clip=true]{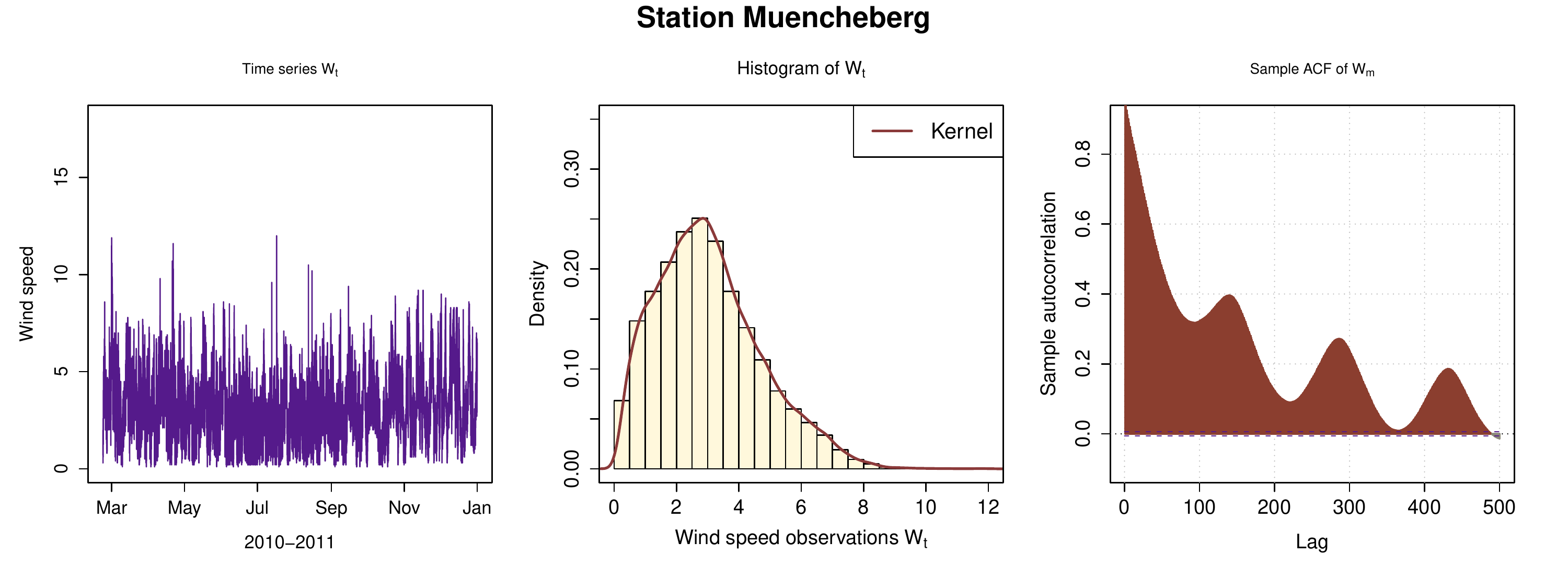}
  \caption{Wind speed data recorded at a 10-minute frequency, corresponding histogram and plot of the autocorrelation function.}\label{figure:stat}
\end{figure}

\begin{figure}[h]
\includegraphics[width=.5\textwidth]{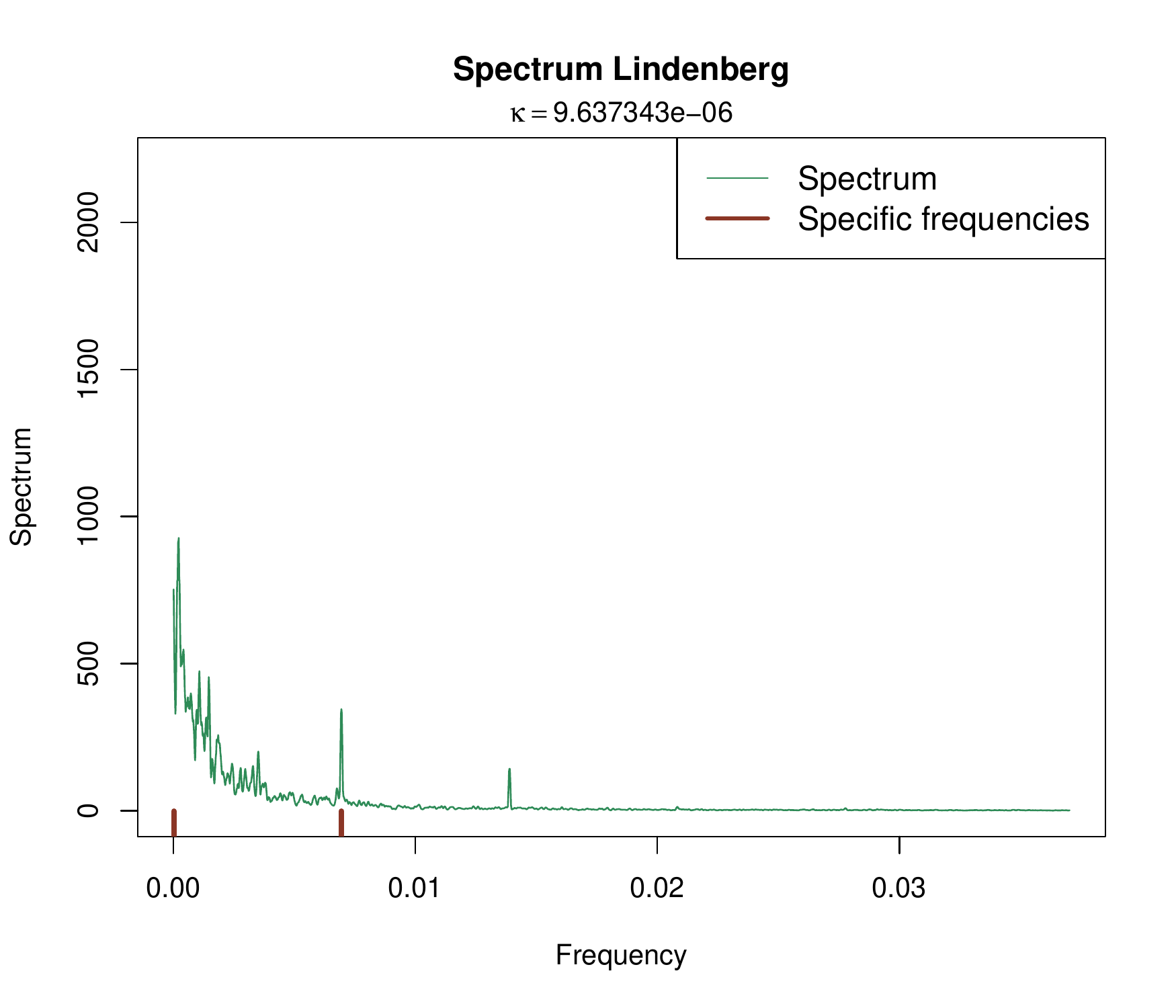}
\includegraphics[width=.5\textwidth]{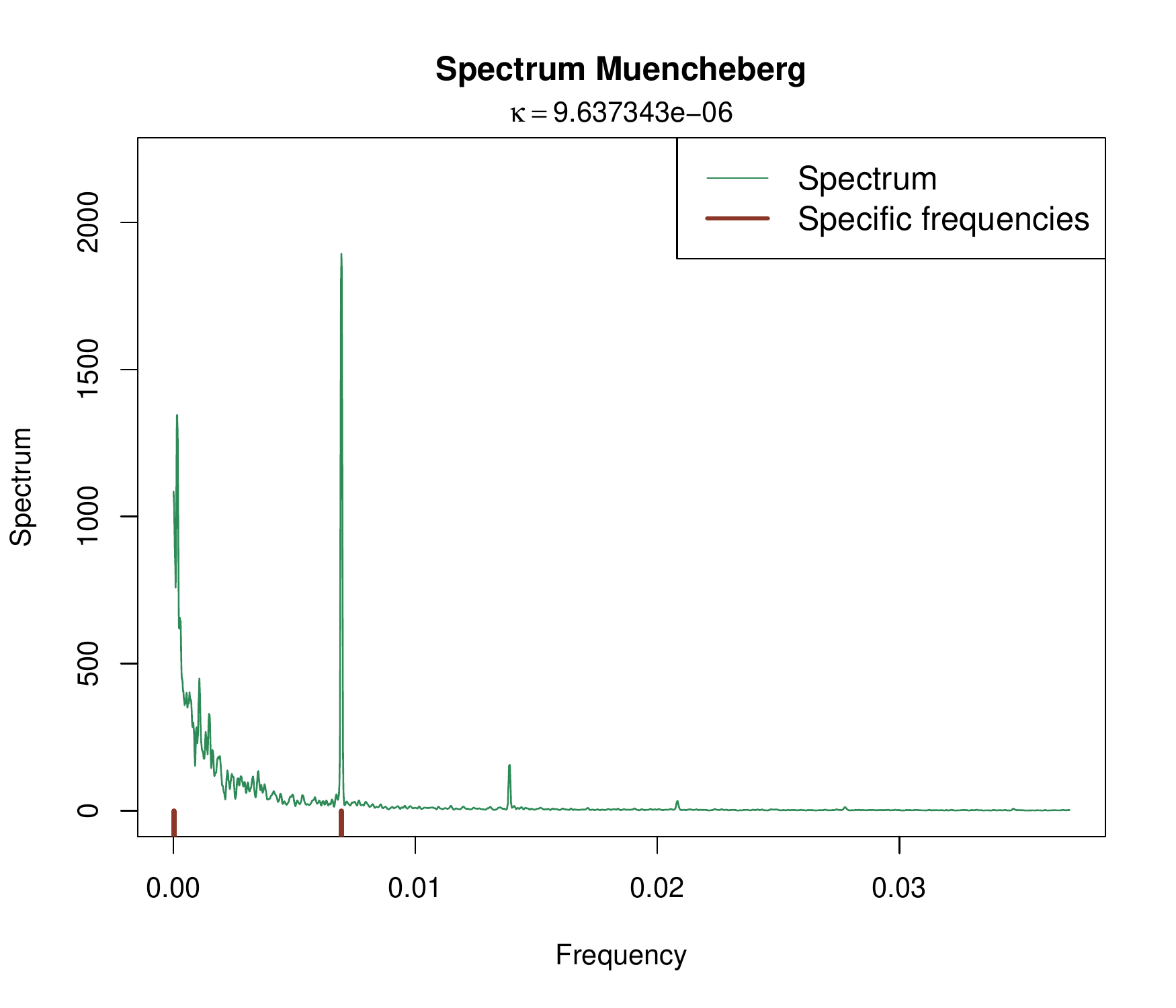}
\caption{Estimation of the spectral density.}
\label{fig:spectrum}
\end{figure}

\begin{figure}[h]
  \fbox{\includegraphics[width=.5\textwidth,trim= .1cm .1cm .1cm .1cm,clip=true]{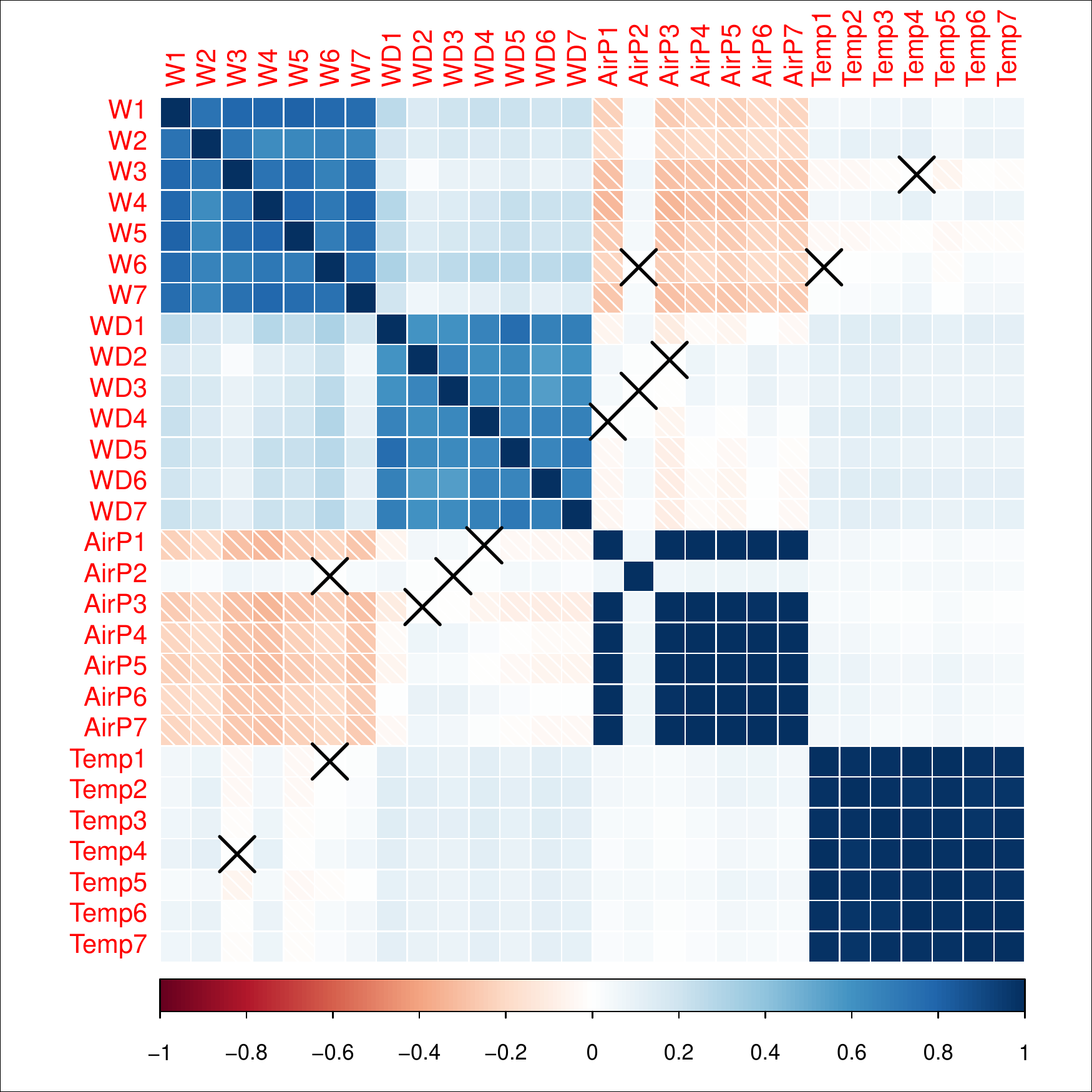}}
  \caption{Plot of pairwise correlation for all dependent and independent variables and all stations, using plain sensor data.}\label{figure:corplot}
\end{figure}

The spatial area of investigated wind speed data is shown in Figure
\ref{figure:brand}. In this article we focus on the
most central station Lindenberg [5]. 
We use M\"uncheberg [2] 
as a reference station and for spatial
information, also use its neighbours.
 They are situated in
Eastern Germany in a region of rural plains. This region is
perfect for wind parks. The data is measured by the ``Deutscher
Wetterdienst'' (DWD) and reaches from January 2009 to December 2011.
For model fitting, a time frame of two and a half years is used and
the remaining months (July 2011 to December 2011) are used for
out-of-sample forecasts. The wind speed $(W_{m,t})_{m\in
\{1,\ldots,M\},t\in\{1,\ldots,T\}}$ is measured in $m/s$ in a 10-minute interval for a station $m$ at time $t$.
\noindent Figure \ref{figure:stat} shows the wind speed data, the corresponding histogram and the autocorrelation function (ACF)  for stations Lindenberg and M\"uncheberg. The observed wind speed possesses strong (conditional) volatility and the histogram of the wind speed shows a non-negative and positively skewed distribution. Moreover, we observe the presence of autocorrelation, which is related to the high frequency of the wind speed data.
\noindent The autocorrelation function shows periodic characteristics. To investigate the periods, we calculate the smoothed periodogram, which is given in Figure \ref{fig:spectrum}. The peaks in this picture depict a periodic behavior for several frequencies, which correspond to diurnal $s_1 =1/0.000019= 52,560$ and annual $s_2 =1/0.00694 = 144$ periods (note that at a data frequency of 10 minutes, there are 144 obs./day and 52,560 obs./year). We observe autocorrelation, a periodic behavior and heteroscedasticity, just as \cite{Ewing2006}, \cite{taylor2009wind} and \cite{ambach2015periodic}.
\noindent According to the fact that wind speed is a spatial
phenomenon, it is reasonable to include the available and relevant
information of neighbouring measurement stations. We will emphasize
this idea with an example. If the wind is blowing from the north to
the south and we observe this information at station M\"uncheberg, we
are able to use this information for the station Lindenberg (cf. Figure \ref{figure:brand}). If the
wind comes from the south, we can consider the observations from
Cottbus. The investigated area of our data set contains seven
different stations and Lindenberg is the midpoint. \\
\noindent Figure \ref{figure:corplot} presents the pairwise correlation of all variables of all stations for the entire in-sample time frame. Blue denotes positive correlation, red denotes negative correlation, darker colors indicate greater values. Crossed boxes represent insignificant correlation coefficients. It can be seen that aside from six exceptions, all correlation coefficients are significant, but some are close to zero. M\"uncheberg's air pressure is barely correlated to any of the other variables. Each station's temperature data is strongly positively correlated to each other station's temperature data. The same holds for the wind direction.
The correlation between most groups of different types of variables is weak. Temperature and wind speed are weakly negatively correlated, wind speed and air pressure are somewhat stronger positively correlated. Eventually, we observe a huge positive correlation between the wind direction and the wind speed, which will be captured by our new multivariate model.
\begin{figure}[h]
  \includegraphics[width=.5\textwidth,trim= .4cm 3.5cm .4cm 3.5cm,clip=true]{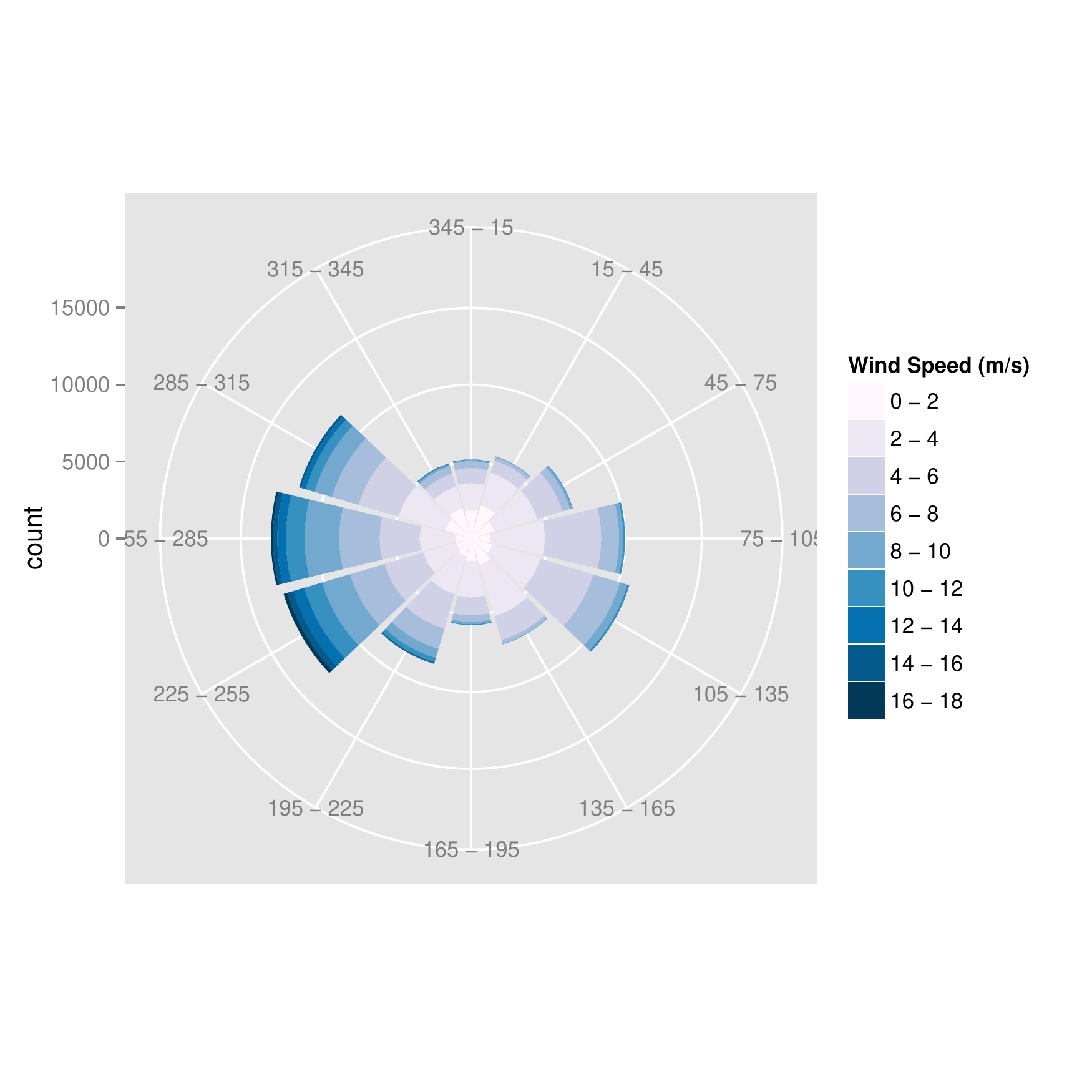}
  \includegraphics[width=.5\textwidth,trim= .4cm 3.5cm .4cm 3.5cm,clip=true]{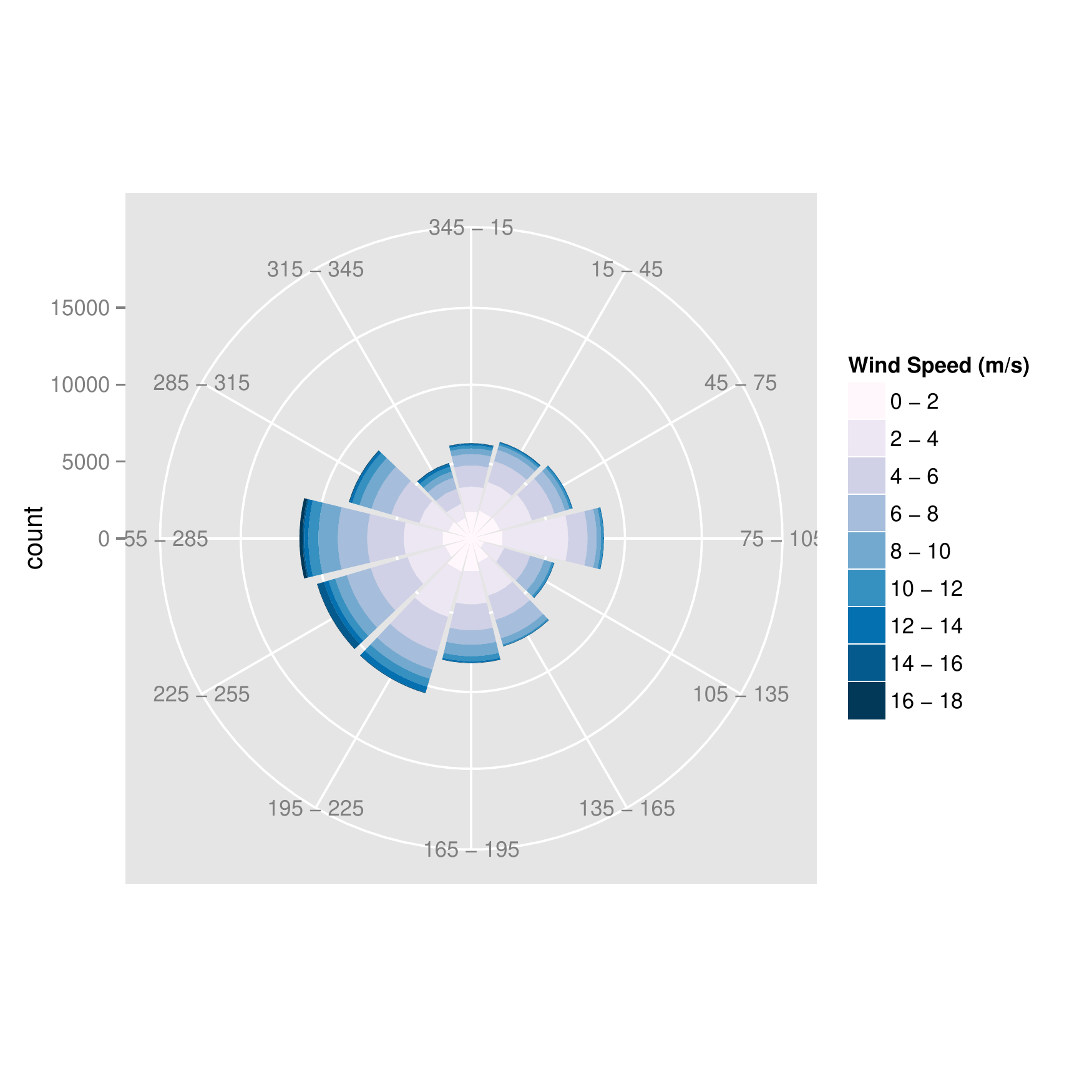}
  \caption{Wind rose, where the wind speed frequencies are plotted by wind direction for station Lindenberg (left) and station M\"uncheberg (right).}\label{figure:rose5}
\end{figure}
\noindent Figure \ref{figure:rose5} depicts the wind rose for our target station Lindenberg (left panel). Wind roses show the densities of wind speeds, dependent on the wind direction. As the wind speed densities are rather large for eastern, north-eastern and south-eastern directions, we conclude that Lindenberg is mostly affected by winds coming from directions at which Baruth, Doberlug-Kirchhain and possibly Berlin-Sch\"onefeld are located. Accordingly, Figure \ref{figure:rose5} shows the wind rose for station M\"un\-che\-berg on the right panel.

\subsection{Benchmark models}\label{section:Benchmarks}

\begin{figure}[h]
\includegraphics[width=.5\textwidth ,trim= .1cm .5cm .1cm 2cm,clip=true]{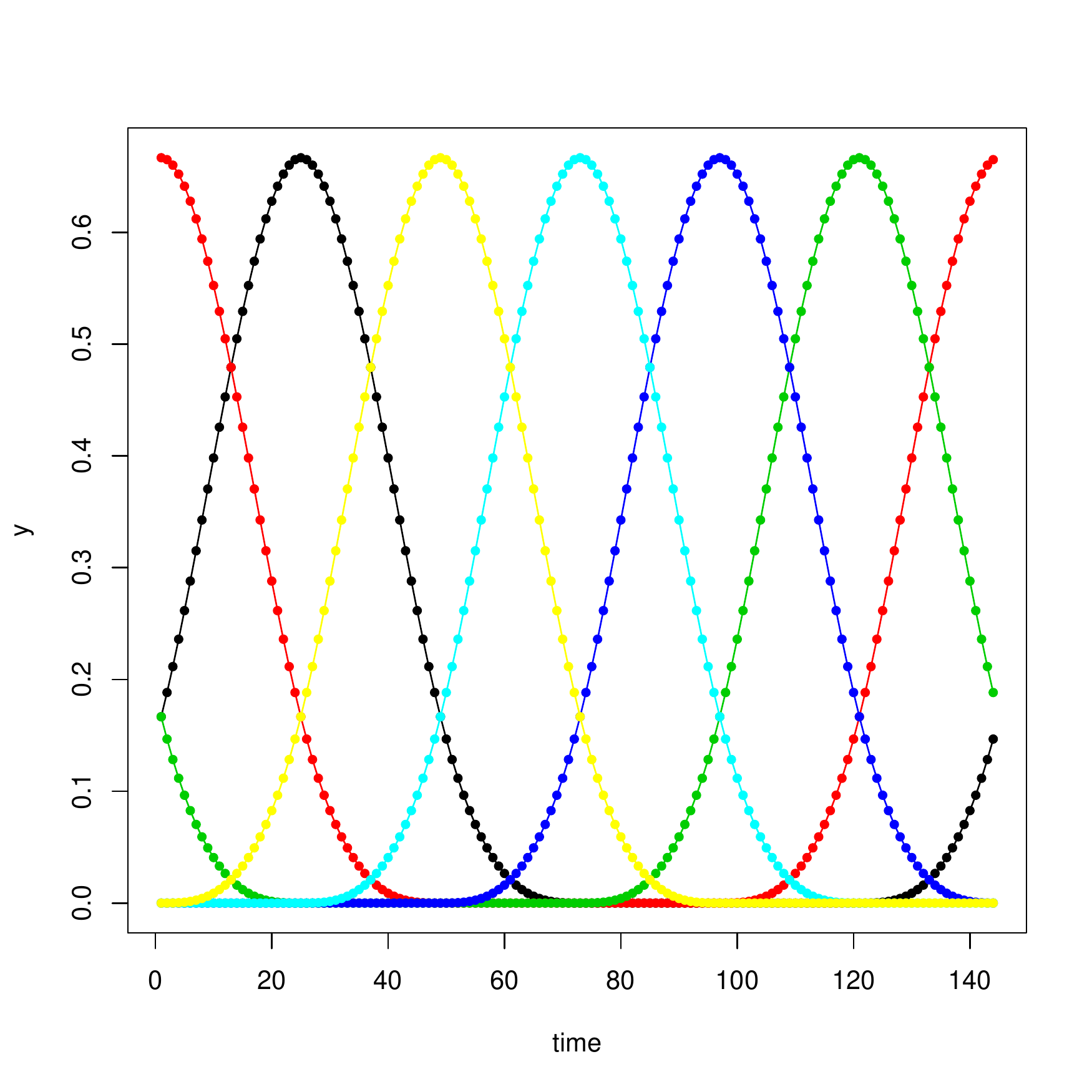}
\caption{Diurnal cubic B-splines.}
\label{fig:splines}
\end{figure}

Univariate wind speed models are based on a decomposition of the time series into a time-dependent intercept, a trend component and a seasonal part. A general model can be

\begin{equation}
\label{general}
    W_{t} = Intercept_{t} + Trend_{t} + Seasonal_{t} + \epsilon_{t},
\end{equation}

\noindent where $W_{t}$ represents the wind speed and $\epsilon_{t}$ is the residual process with zero mean and time dependent variance $\sigma_{t}$ for time $t$. One very general model of this type is the ARFIMA-APARCH approach with periodic regressors as proposed by \cite{Ambach2014}, which nests various other models formerly proposed. As we contribute a novel multivariate model, we use the ARFIMA-APARCH merely as a competitive benchmark from the class of univariate models. As in the original motivation of this approach, we include periodic B-splines instead of Fourier series to capture periodicity/seasonality. The intercept and the seasonal components are modelled by
\begin{equation}
\label{eq:Model1Seas}
  Intercept_{t} + Seasonal_{t}   = \vartheta_{11} + \sum^{k_1}_{i_1 = 2} \sum^{k_2}_{i_2 = 2} \vartheta_{i_1, i_2} f_{i_1}^{s_1}(t) f_{i_2 }^{s_2}(t).
 \end{equation}
\noindent where $f_{i_{1}}^{s_{1}}(t)$ and $f_{i_{2}}^{s_{2}}(t)$ are up to $k_1$ and $k_2$ time dependent periodic functions. The residual process $\epsilon_{t}$ is stationary and follows a ARFIMA-APARCH. Thus, it is given by
\begin{eqnarray}
\label{equation:arfimaaparch}
\begin{array}{rcl}
\epsilon_{t} &\equiv & \phi (B) (1- B)^{d}  X_{t}  =  \theta (B) Z_{t} ,  \\[.2cm]
 & Z_{t}=& h_{t} \eta_{t}  \qquad \text{where} \quad   \eta_{t} \overset{iid}{\sim} WN(0, \sigma^2)  \\[.2cm]
       &h_{t}^\delta =&    \alpha_{0} + \sum\limits_{l=1}^{Q} \alpha_l (| Z_{t-l}| - \gamma_l Z_{t-l})^{\delta}  + \sum\limits_{i=1}^{P} \beta_i h_{t-i}^{\delta} , \vspace*{0.3cm} \vspace*{0.4cm}
 \end{array}
\end{eqnarray}
where $d \in (-0.5, 0.5)$ is the differencing parameter and $B$ is the backward shift operator with $B^u X_{t} = X_{t - u}$. The ARMA(p,q) polynomials $\phi$ and $\theta$ are $\phi (B) = 1- \phi_1 B - \ldots - \phi_p B^p $ and $\theta (B) = 1+ \theta_1 B + \ldots + \theta_q B^q$ and have no common factors. The roots of $\phi$ and $\theta$ must lie outside the unit circle. \cite{Ding1993} point out that instead of the conditional variance, the conditional standard deviation with power $\delta$ follows an asymmetric power ARCH (APARCH) process. The asymmetry parameter is $\gamma_l \in [-1, 1]$ for $l = 1,...,Q$. \cite{Ding1993} provide the existence of a stationary solution for the APARCH process by $\sum\nolimits_{l=1}^{Q} \alpha_l E(|z| - \gamma_l z)^{\delta}  + \sum\nolimits_{i=1}^{P} \beta_i  < 1$.\\
The parameters of this ARFIMA-APARCH model are usually estimated by means of a quasi maximum likelihood (QML) estimation procedure, which can be slow, requires a strict distributional assumption and is prone to the usual problems of numeric solving, e.g. identification of the global maximum.\\
\cite{Ambach2014} describes seasonality modeling by means of periodic B-splines, which are becoming more and more accepted in the literature. \cite{Thapar2011}, \cite{Bazilevs2012} and \cite{LeGuyader2014} show the power of this approach. The fundamental basics of the spline functions are defined by \cite{de1978practical} and \cite{eilers1996flexible}. In the application, we define our B-Splines by a set of equidistant knots such that each function oscillates once per day. Also, we use a second setup of functions to capture annual periodicity. All B-Splines are twice continuously differentiable. Figure \ref{fig:splines} depicts our diurnal functions.\\
\noindent Aside from the nested ARFIMA-APARCH model for $\epsilon_{t}$, we consider a simple $VAR(p)$ (vector autoregressive) and an $AR(p)$ process. The vector contains wind speed, but also several other variables, which are described in Section \ref{section:multivariate}. Finally, we investigate the usual persistence or ``na\"ive'' benchmark \citep[see][]{Nielsen1998}. The na\"ive predictor uses the current value as a forecast for future values $\widehat{W}_{t + k} = W_t$.

\subsection{Novel multivariate wind speed prediction
model}\label{section:multivariate}

\begin{figure}
  \begin{center}
\begin{tikzpicture}
[auto,
decision/.style={diamond, draw=gray, thick, fill=blue!10, text width=5em, text badly centered, inner sep=1pt},
block/.style ={rectangle, draw=gray, thick, fill=blue!10, text width=30em, text centered, rounded corners, minimum height=4em},
line/.style ={draw, thick, -latex',shorten >=0pt},
cloud/.style ={draw=red, thick, ellipse, fill=red!20, minimum height=4em}]
\matrix [column sep=13mm,row sep=5mm]
{
& \node [block] (1) {(1) Set an initial $1 \times T$ weight $\boldsymbol{\omega} = (1,\ldots,1)$\\ and set the iteration parameter to $\mathcal{K} =1$}; & \coordinate (dummy-write);\\
& \node [block] (2) {(2) Estimate \eqref{eq:SVAR} using either lasso or elastic net with weights  $\boldsymbol{\omega} $}; & \coordinate (dummy-write-done);\\
& \node [block] (3) {(3) Estimate $\boldsymbol{{\sigma}_t}$ by \eqref{eq:TARCH} using again the lasso or elastic net and $\{|\boldsymbol{\widehat{\epsilon_t}}|\}$}; & \\
& \node [block] (4) {(4) Subsequently, redefine the vector $\boldsymbol{\omega}$ by $\boldsymbol{\omega} = \left(\left(\boldsymbol{\widehat{\sigma}_{1}^{-2}}\right),...,\left(\boldsymbol{\widehat{\sigma}_{T}^{-2}}\right) \right)$ \\ and use the values obtained by (3).}; & \coordinate (dummy5);\\
& \node [block] (5) {(5) The convergence of $\boldsymbol{{\sigma}_t}$ is used as abort criterion if
$\boldsymbol{\Delta_{\mathcal{K}}} = || \widehat{\sigma}_{\mathcal{K}-1}-\widehat{\sigma}_{\mathcal{K}} || < 10^{-3}$. Convergence attained?}; & \coordinate (dummy-review-done);\\
& \node [block] (6) {Stop the algorithm.}; & \\
};
\begin{scope}[every path/.style=line, rounded corners]
\path (1) -- (2);
\path (2) -- node [midway] {} (3);
\path (3) -- (4);
\path (4) -- (5);
\path (5.east) -- node[above] {No} (dummy-review-done) -- (dummy-write-done)  -- node[above] {$\mathcal{K} = \mathcal{K} +1$}(2.east);
\path (5) -- node [midway] {Yes} (6);
\end{scope}
\end{tikzpicture}
  \end{center}
  \caption{Estimation scheme.}\label{fig:scheme}
\end{figure}
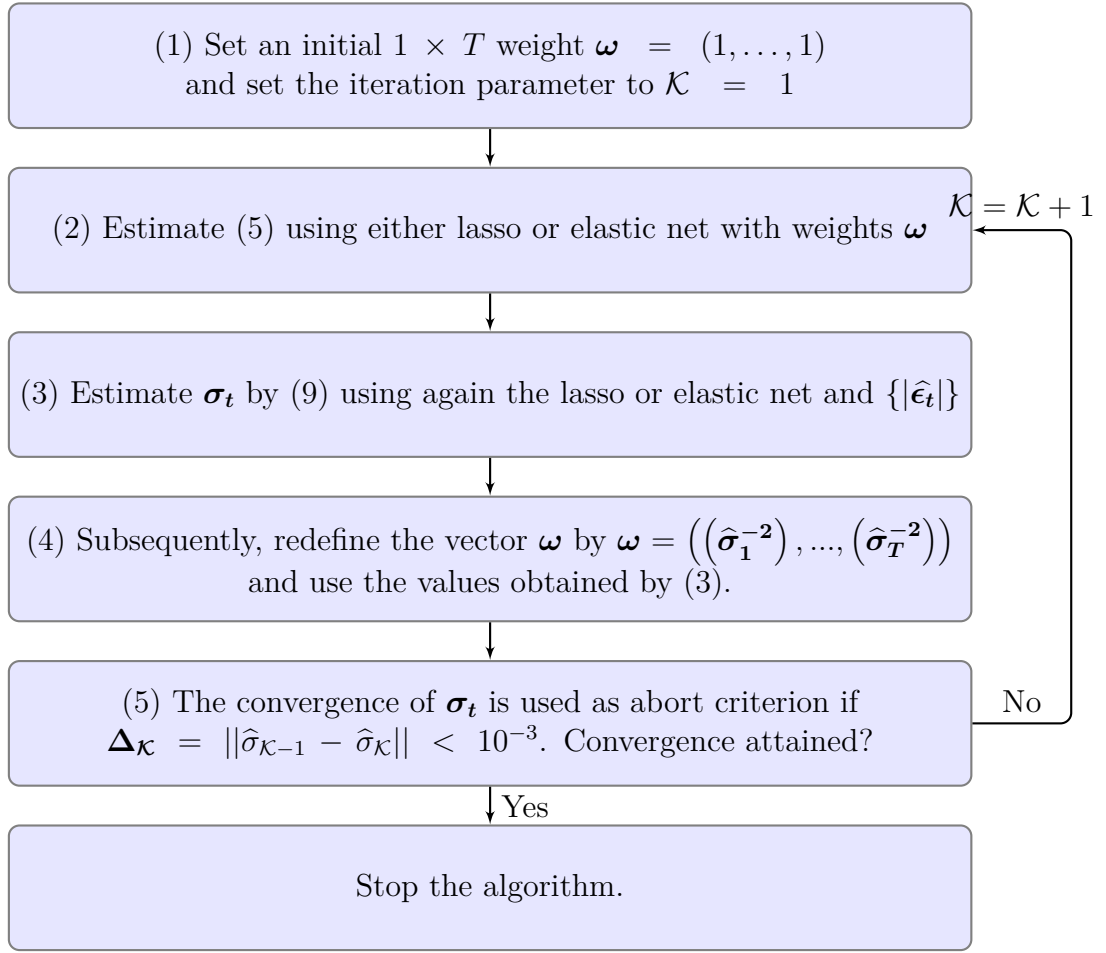

In contrast to the established univariate modeling by, e.g., the above-mentioned ARFIMA-APARCH process, we introduce a multivariate model. Basically, our model is a VAR, but the mean is enriched by seasonal functions and the conditional variance of the residual process is modelled by a threshold autoregressive process (TARCH). The mean part as well as the variance part may contain external periodic regressors (X), so the model can be described as an SVARX-TARCHX approach.\\
We decompose our wind direction information into an east-west component, given by $\sin (az_{m,t})$, and a north-south component, given by $\cos (az_{m,t})$, where $az_{m,t}$ is the azimuth, as measured at station $m$ for time $t$. The SVARX part captures wind speed information, but also, air pressure information $AP_{m,t}$ and temperature $C_{m,t}$, measured in degrees Celsius are incorporated. For stations $m = 1, \ldots, M$ and time point $t = 1, \ldots, T$, the dependent vector consists of observations for 5 variables, i.e. $\boldsymbol{Y}_t \in \mathbb{R}^{5\cdot M}$, which is
\begin{equation}\label{eq:1}
\begin{array}{lr}
\boldsymbol{Y_t} = & \bigg(
W_{1,t}, \; \ldots, \; W_{M,t}, \; \; \; \sin(az_{1,t}), \; \ldots, \; \sin(az_{M,t}),  \\
 & \cos(az_{1,t}), \; \ldots, \; \cos(az_{M,t}),  \; \; \;  AP_{1,t}, \; \ldots, \;  AP_{M,t}, \; \; \; C_{1,t}, \; \ldots, \; C_{M,t}
\bigg)'.
\end{array}
\end{equation}
\noindent The model is accounting for an overall relationship between speed and direction variables across all sites. The multivariate dependent variables in $\boldsymbol{Y_t}$ are described by a vector autoregressive model for the mean part given by
\begin{eqnarray}\label{eq:SVAR}
                    \boldsymbol{Y_t} &=& \boldsymbol{\vartheta}_t  
                     + \sum_{j=1}^{{J}} \boldsymbol{\phi}_{t,j} \boldsymbol{Y}_{t-j} + \boldsymbol{\epsilon}_t ,\\
                    \boldsymbol{\vartheta}_t &=& \boldsymbol{\vartheta}_{0}
                    + \sum^{k_1}_{i_1=2} \boldsymbol{\vartheta}_{i_1,1} {f}_{i_1}^{s_1}(t)
                    + \sum^{k_2}_{i_2=2} \boldsymbol{\vartheta}_{1, i_2} {f}_{i_2}^{s_2}(t)
                    + \sum^{k_1}_{i_1=2} \sum^{k_2}_{i_2=2} \boldsymbol{\vartheta}_{i_1, i_2} {f}_{i_1}^{s_1}(t) {f}_{i_2}^{s_2}(t), \\
                    \boldsymbol{\phi}_{t,j}  &=&  \boldsymbol{\phi}_{0,j}
                    + \sum^{k_1}_{i_1=2} \boldsymbol{\phi}_{i_1,1,j} {f}_{i_1}^{s_1}(t)
                    + \sum^{k_2}_{i_2=2} \boldsymbol{\phi}_{1, i_2,j} {f}_{i_2}^{s_2}(t)
                    + \sum^{k_1}_{i_1=2} \sum^{k_2}_{i_2=2} \boldsymbol{\phi}_{i_1, i_2, j} {f}_{i_1}^{s_1}(t) {f}_{i_2}^{s_2}(t), \\
                    \boldsymbol{\epsilon}_t &=&  \boldsymbol{\sigma}_t \boldsymbol{\eta}_t ,
\end{eqnarray}
\noindent where $\{\boldsymbol{\eta}_t\}$ is i.i.d., $\mathrm{E}(\boldsymbol{\eta}_t) = \boldsymbol{0}$ and $\mathrm{Var}(\boldsymbol{\eta}_t) = \boldsymbol{1}$. Moreover, $\boldsymbol{\vartheta}_0$ is an $(M \cdot 5) \times 1$ intercept vector and $\boldsymbol{\vartheta}_{i_1, 1}$, $\boldsymbol{\vartheta_{1, i_2}}$ and $\boldsymbol{\vartheta_{i_1, i_2}}$ are $(M \cdot 5) \times 1$ periodic coefficient vectors. $\boldsymbol{\phi}_{0,j}$, $\boldsymbol{\phi}_{i_1,1,j}$, $\boldsymbol{\phi}_{1,i_2,j}$ and $\boldsymbol{\phi}_{i_1,i_2,j}$ are $(M \cdot 5) \times (M \cdot 5)$ parameter matrices of autoregressive and periodic autoregressive parameters for lag $j \in \mathbb{N}$. The periodic B-spline functions ${f}_{i_1}^{s_1}(t)$ and ${f}_{i_2}^{s_2}(t)$ are similar to Section \ref{section:Benchmarks}. Finally, $\boldsymbol{\sigma}_t$ and $\boldsymbol{\eta}_t$ are $(M \cdot 5) \times 1$ vectors which follow a TARCH process \citep[see][]{GJR1993} given by
\begin{eqnarray}\label{eq:TARCH}
    \boldsymbol{\sigma}_t &=& \boldsymbol{\alpha}_{0} 
    + \sum_{i_1=2}^{k_1} \boldsymbol{\alpha}_{i_1} {f}_{i_1}^{s_1}(t) 
    + \sum_{h=1}^{{P}} \boldsymbol{\zeta}_{t,h} \boldsymbol{I}_t^{+} \boldsymbol{\epsilon}_{t-h} 
    + \sum_{l=1}^{{Q}} \boldsymbol{\psi}_{t,l} \boldsymbol{I}_t^{-} \boldsymbol{\epsilon}_{t-l} ,\\
\boldsymbol{\zeta}_{t,h} &=&  \boldsymbol{ \zeta}_{0, h}
 + \sum^{k_1}_{i_1=2} \boldsymbol{ \zeta}_{i_1, h} {f}_{i_1}^{s_1}(t) , \\
\boldsymbol{\psi}_{t,l} &=&  \boldsymbol{ \psi}_{0, l} 
+ \sum^{k_1}_{i_1=2} \boldsymbol{ \psi}_{i_1, l} {f}_{i_1}^{s_1}(t) ,
\end{eqnarray}
\noindent where $\boldsymbol{\alpha}_{0}$ and $\boldsymbol{\alpha}_{i_1}$ are $(M \cdot 5) \times 1$ parameter vectors. Furthermore, $\boldsymbol{\zeta}_{0,h}$, $\boldsymbol{\psi}_{0,l}$, $\boldsymbol{\zeta}_{i_1,h}$ and $\boldsymbol{\psi}_{i_1,l}$ are autoregressive and periodic autoregressive parameter matrices ($(M \cdot 5) \times (M \cdot 5)$) within the variance. Hence, the $M \cdot 5 \times 1$ vectors of indicator functions $I_t^{+}$ and $I_t^{-}$ are given by
\begin{eqnarray}
\begin{array}[t]{lcr}
 I_{\mathcal{N},m,t-1}^{+} = \begin{cases} 1 , 
 \qquad
 \epsilon_{\mathcal{N},m,t-1} > 0 \\ 0, 
  \qquad
 \epsilon_{\mathcal{N},m,t-1} \leq 0 \end{cases} & \text{,}
&  I_{\mathcal{N},m,t-1}^{-} = \begin{cases} 1, 
\qquad
\epsilon_{\mathcal{N},m,t-1} \leq 0 \\ 0, 
\qquad
\epsilon_{\mathcal{N},m,t-1} > 0 \end{cases},
\end{array}
\end{eqnarray}
\noindent where $\mathcal{N} \in \{1,...,5\}$ represents the $\mathcal{N}$th regressor. These vectors provide one way to model a TARCH process for a station $m$. Therefore, we double the parameter space, but are able to differentiate between negative and positive shocks. After all, the number of parameters amounts to
\begin{equation*}
2\cdot\left(J + 2\left(\left(k_1 - 1\right) + \left(k_2 - 1\right) + \left(k_1 k_2 - 1\right) + 1\right) + \left(k_1 - 1\right) + P + Q + 1 + 2k_1 \right),
\end{equation*}
which may be much larger than the number of observations. Thus, regularization and shrinkage are necessary.\\
\cite{mbamalu1993load} use an iteratively re-weighted least squares approach to develop accurate load forecasts. More recently, \cite{ziel2014} implement an iteratively re-weighted lasso method to predict the electricity price. We take the lasso and the elastic net as estimation methods into consideration, as they are designed to reduce the parameter space in the estimation stage. \cite{ren2010} take the lasso method in the context of vector autoregressive models. The weighted lasso estimation of the parameter vector $\boldsymbol{\theta_{\mathcal{N},m}}$ for both the SVARX model \eqref{eq:SVAR} and the TARCHX model \eqref{eq:TARCH} is given by
\begin{equation}\label{eq:lasso}
\boldsymbol{\hat{\theta}_{\mathcal{N},m}} = \underset{{\tiny{ \boldsymbol{\theta_{\mathcal{N},m}}}} \in \mathbb{R}^{p_j}} {\arg \min} \sum_{t=1}^T({Y_{\mathcal{N},m,t}} - {\omega_t} \boldsymbol{X_{\mathcal{N},m,t}} \boldsymbol{{\theta_{\mathcal{N},m,t}}} )^2 + \lambda_{\mathcal{N},m,T} \sum_{j = 1}^{p_j} |\boldsymbol{\theta_j}|,
\end{equation}
where $\omega_t$ is the weight vector $\boldsymbol{\omega} = (\omega_1,...,\omega_T)$, $p_j$ being the number of elements of the parameter vector $\boldsymbol{\theta_{\mathcal{N},m}}$ and the tuning parameter is $\lambda_{\mathcal{N},m,T} \geq 0$. The second estimation method is the elastic net \citep[see][]{friedman2009elements} which is given by
\begin{eqnarray}\label{eq:elnet}
\begin{array}[t]{ll}
\boldsymbol{\hat{\theta}_{\mathcal{N},m}} = \underset{{\tiny{ \boldsymbol{\theta_{\mathcal{N},m}}}} \in \mathbb{R}^{p_j}} {\arg \min} \sum_{t=1}^T({Y_{\mathcal{N},m,t}} - {\omega_t} \boldsymbol{X_{\mathcal{N},m,t}} \boldsymbol{{\theta_{\mathcal{N},m,t}}} )^2 &+ \lambda_{\mathcal{N},m,T} \alpha \sum_{j = 1}^{p_j} |\boldsymbol{\theta_j}| \\ &+ \frac{1}{2}\lambda_{\mathcal{N},m,T} (1- \alpha )\sum_{j = 1}^{p_j} \boldsymbol{\theta_j}^2,
\end{array}
\end{eqnarray}
where $\alpha \in [0,1]$ provides an additional tuning parameter. In the case of $\alpha = 1$, we obtain the lasso case and if $\alpha = 0$, we obtain the ridge regression. Finally, if $\alpha \in (0,1)$, the penalties represent the elastic net.\\
The estimation is done in two-steps. We start with the estimation of the mean model and afterwards, we re-weight the mean model using $\boldsymbol{\widehat{\sigma}_t}$. Figure \ref{fig:scheme} provides the estimation scheme for the iteratively re-weighted lasso method \citep[see][]{efron2004,ziel2014}. During the algorithm, we use the Akaike information criterion (AIC) for the model selection.
The procedure is repeated until some convergence is achieved. If $\mathcal{K} = 1$, we obtain the homoscedastic estimates without re-weighting the lasso. The advantage of this approach is the fast computing time, and further extensions are still possible. Clearly, we have to choose the specific lags of the AR and the TARCH part, which we discuss in the next section.\\
From the modeling perspective, we expect that the iteratively re-weighted lasso or elastic net method should be superior compared to the ML/QML approaches, because we are able to handle a huge parameter space with a variable selection and optimization algorithm. Moreover, we do not need a specific distributional assumption.

\section{Model fitting results}\label{section:insample}

\begin{figure}[h]
  \includegraphics[width=.5\textwidth,trim= .3cm .8cm .5cm 1.5cm,clip=true]{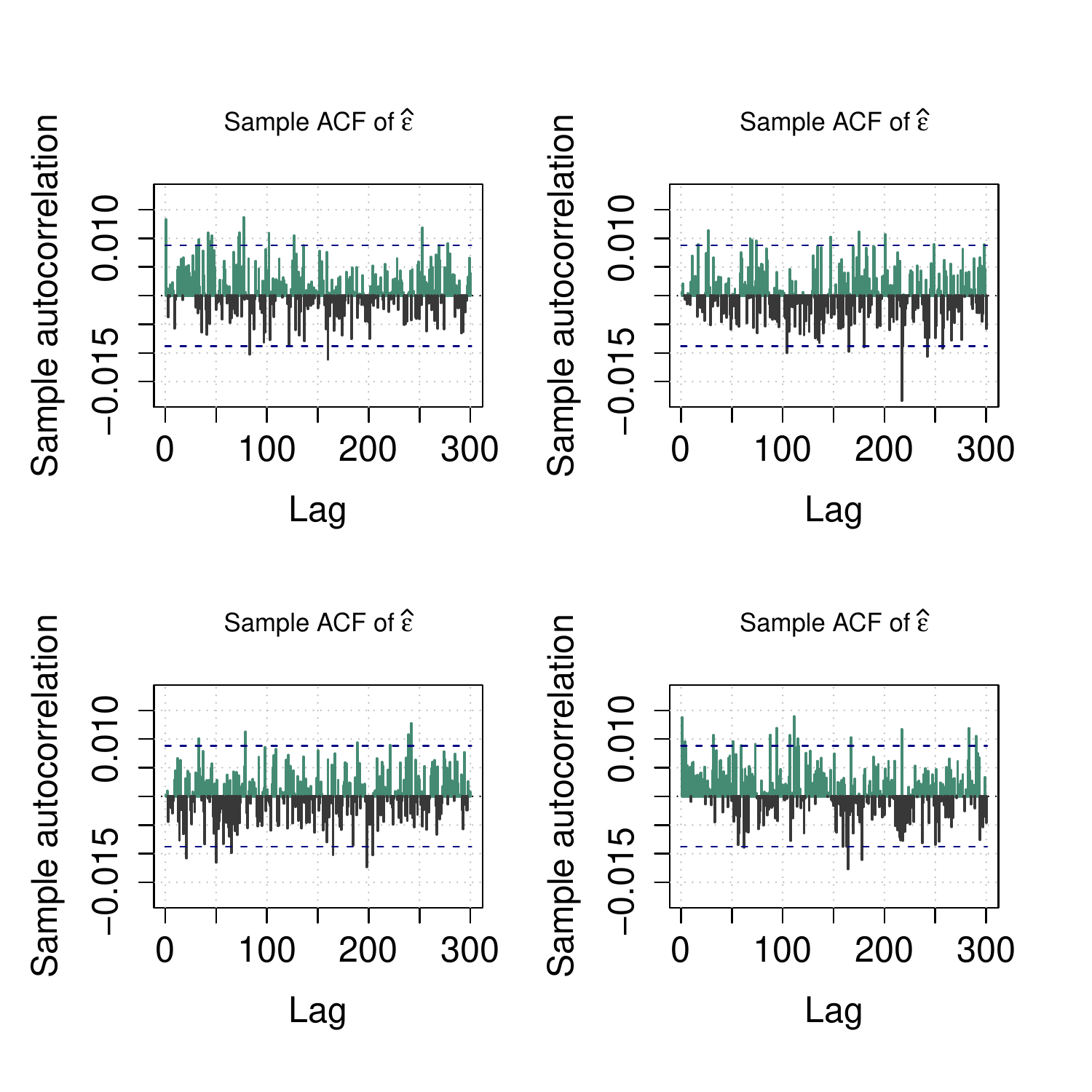}
   \includegraphics[width=.5\textwidth,trim= .3cm .8cm .5cm 1.5cm,clip=true]{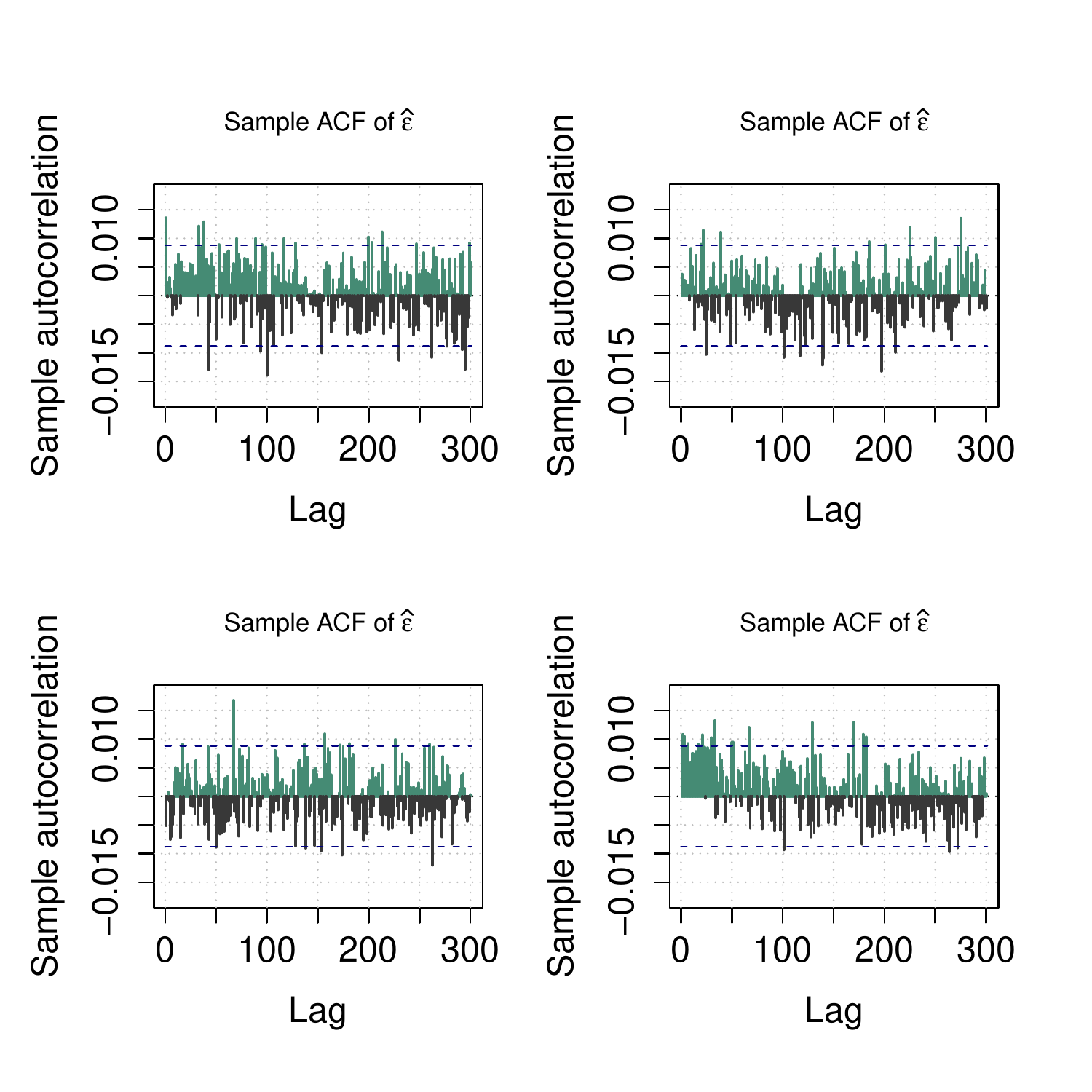}
  \caption{Autocorrelation function (ACF) of $\{\boldsymbol{\widehat{\epsilon}_t}\}$ for the new model with lasso method (first and second column) and ACF of the elastic net (third and forth column).}
  \label{figure:acf}
\end{figure}

\begin{figure}[h]
  \includegraphics[width=.5\textwidth,trim= .3cm .8cm .5cm 1.5cm,clip=true]{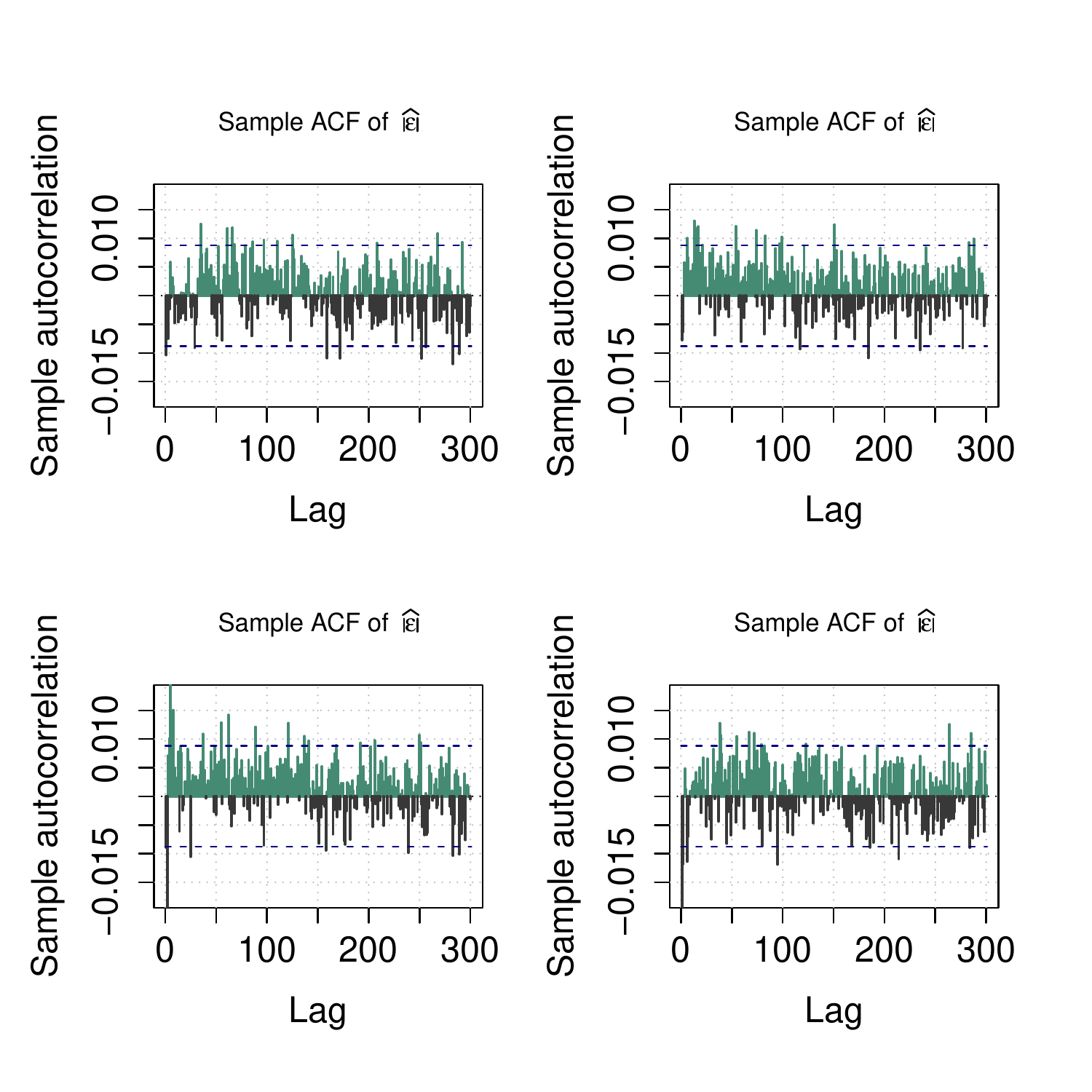}
    \includegraphics[width=.5\textwidth,trim= .3cm .8cm .5cm 1.5cm,clip=true]{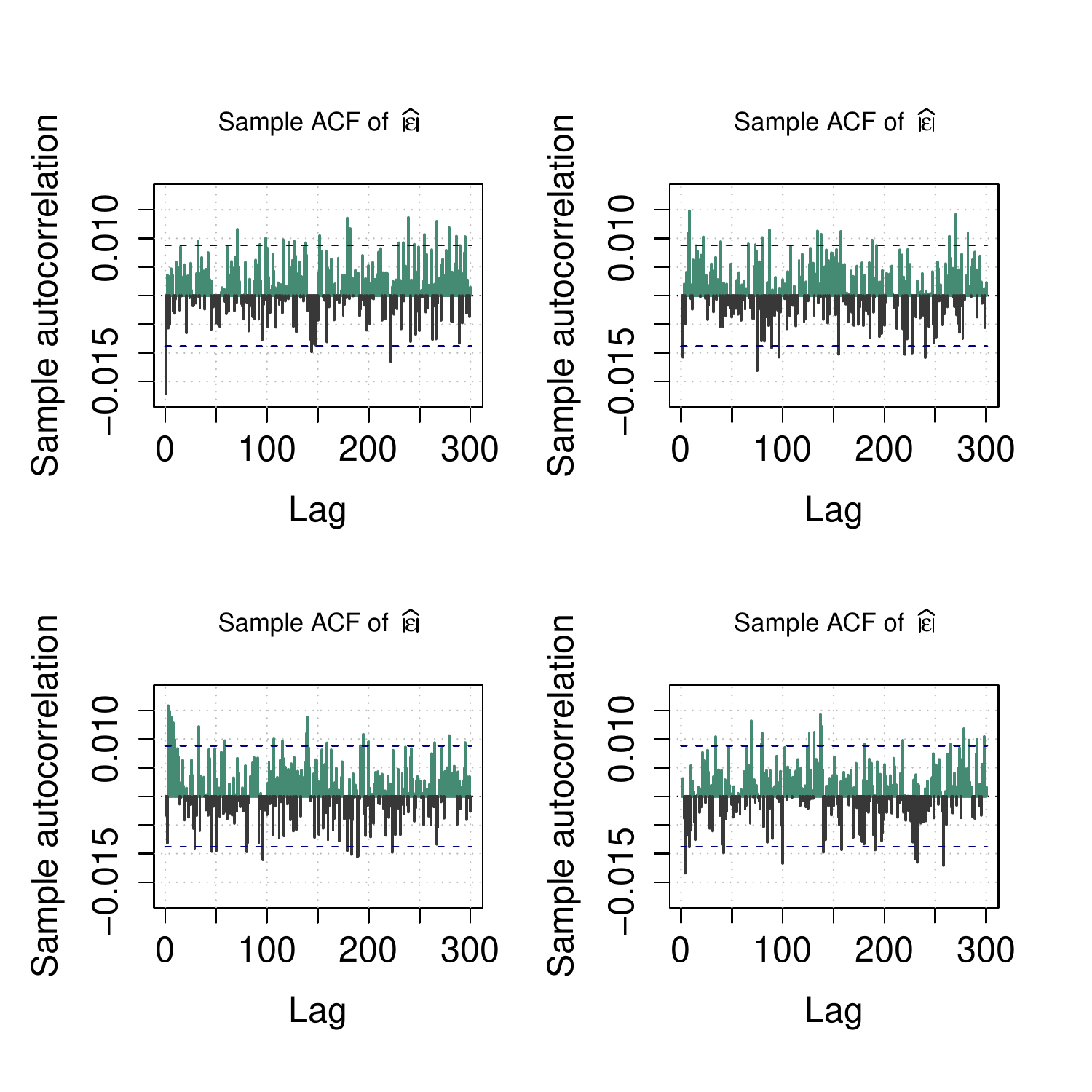}
    \caption{Autocorrelation function of $\{\boldsymbol{|\widehat{\epsilon}_t|}\}$ (ACF) for the new model with lasso method (first and second column) and ACF of the elastic net (third and forth column).}
  \label{figure:absacf}
\end{figure}

The ARFIMA-APARCH process presented in Equations
\eqref{eq:Model1Seas} and \eqref{equation:arfimaaparch} is our extensive benchmark
model. Here, this model is estimated by a QML approach under
normally distributed residuals. According to the complex ARFIMA-APARCH,
we determine a sparse parametrization. The in-sample results are
comparable to previous findings by \cite{Ambach2014} and
\cite{taylor2009wind}. \\
The novel wind speed approach is described in Equations \eqref{eq:SVAR} and \eqref{eq:TARCH}. The estimation of the model is done by an iteratively re-weighted method. We distinguish between two estimation methods, the lasso method \eqref{eq:lasso} and the elastic net \eqref{eq:elnet}. An advantage of these methods is that in contrast to the QML estimation of the ARFIMA-APARCH benchmark model, we do not need a
distributional assumption for the SVARX-TARCHX model.\\
The wind speed time series shows a huge presence of autocorrelation with a
strong diurnal structure. Therefore, we choose to take an
autocorrelation structure of about two days, which is
${J} = 288 + 1$. The TARCH conditional variance structure
of our model includes a periodicity of two days, i.e.
${Q} = {P} = 288 + 1$ lags. The covariance
structure between each station and each response variable is also
modelled by the vector autoregressive lags within mean and variance
part, but some interactions may be set to zero during the lasso/elastic net iterations.\\
Figure \ref{figure:acf} depicts the autocorrelation functions of
station M\"uncheberg and Lindenberg for the standardized residuals
of the wind speed. Each $2\times 2$ block represents the respective angle on the diagonals: The main diagonal (top-left and bottom-right) represents the ACF of stations 2 and 5. The off-diagonal (top-right and bottom-left) represents cross-correlations: Top-right represents the ACF of station 2 on station 5, bottom-left represents ACF of station 5 on station 2. Quite few of them (in all cases, fewer than 5 \%) are significant.\\
\noindent Figure \ref{figure:absacf} depicts the
autocorrelation functions of station M\"uncheberg and Lindenberg for
the absolute standardized residuals of the wind speed series
$\{|\boldsymbol{\widehat{\epsilon}_t|}\}$ and both estimation methods.
The ACFs in Figures \ref{figure:acf} and \ref{figure:absacf} show only a minor presence of remaining autocorrelation, as there are very few single spikes outside the confidence bands.\\
Additionally to ACF plots, we calculate
the Ljung-Box test for $\{\boldsymbol{\widehat{\epsilon_t}}\}$  and
$\{|\boldsymbol{\widehat{\epsilon_t}}|\}$. Applying a level of
significance $5\%$, we cannot reject the null hypothesis of
independence. After all, the autocorrelation analysis suggests an
excellent model fit, especially due to the fact that almost no
periodic structure remains in the residuals. Therefore, we expect
proper forecasting results.

\section{Out-of-sample forecasting results}\label{section:oosample}

\begin{figure}[h]
  \includegraphics[width=1\textwidth, height = 6cm,trim= .01cm .3cm .2cm .7cm,clip=true]{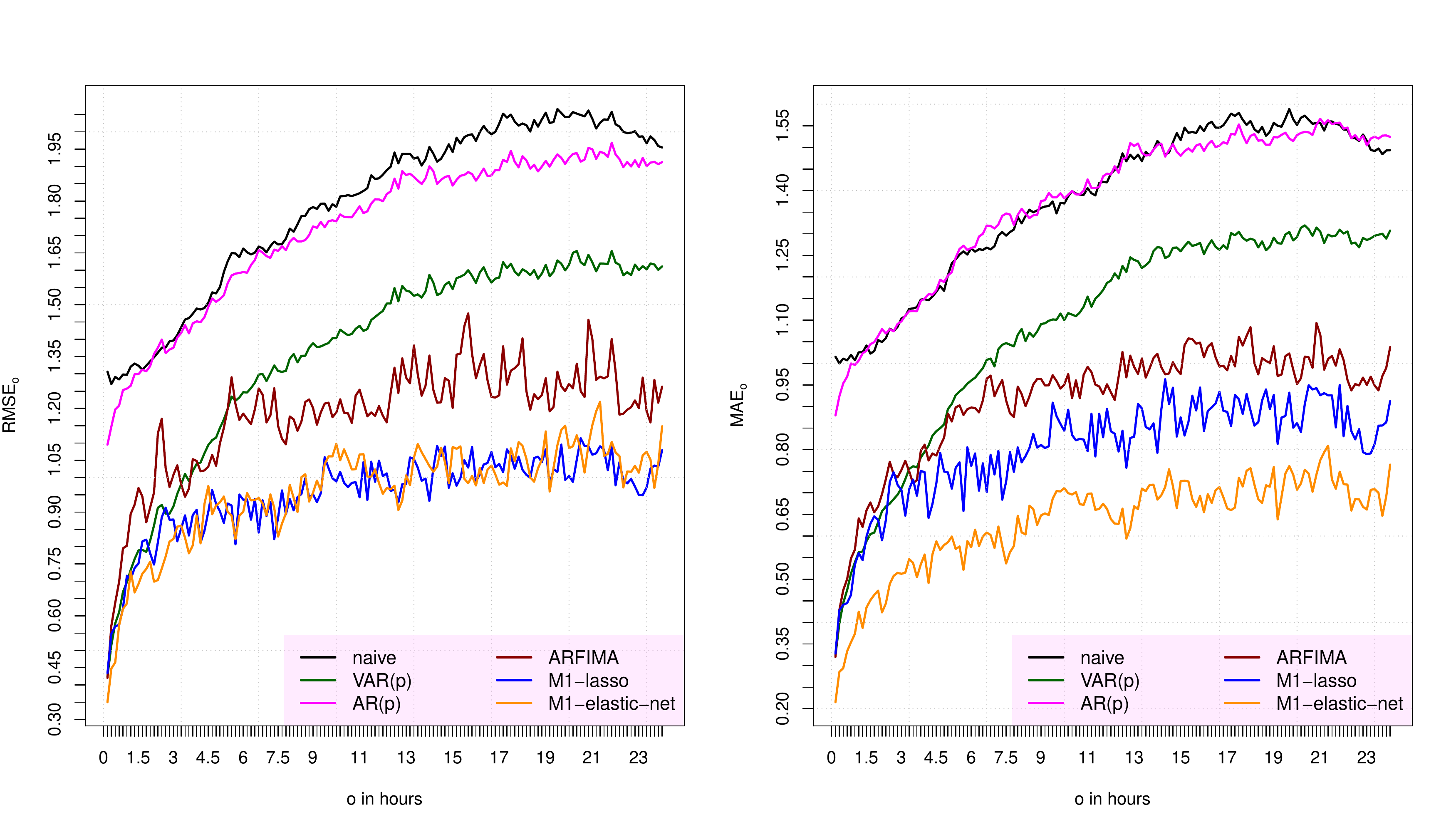}
  \caption{RMSE (first column) and MAE (second column) by forecasting horizon for station M\"uncheberg for all models.}\label{figure:meas1}
  \includegraphics[width=1\textwidth, height = 6cm,trim= .01cm .3cm .2cm .7cm,clip=true]{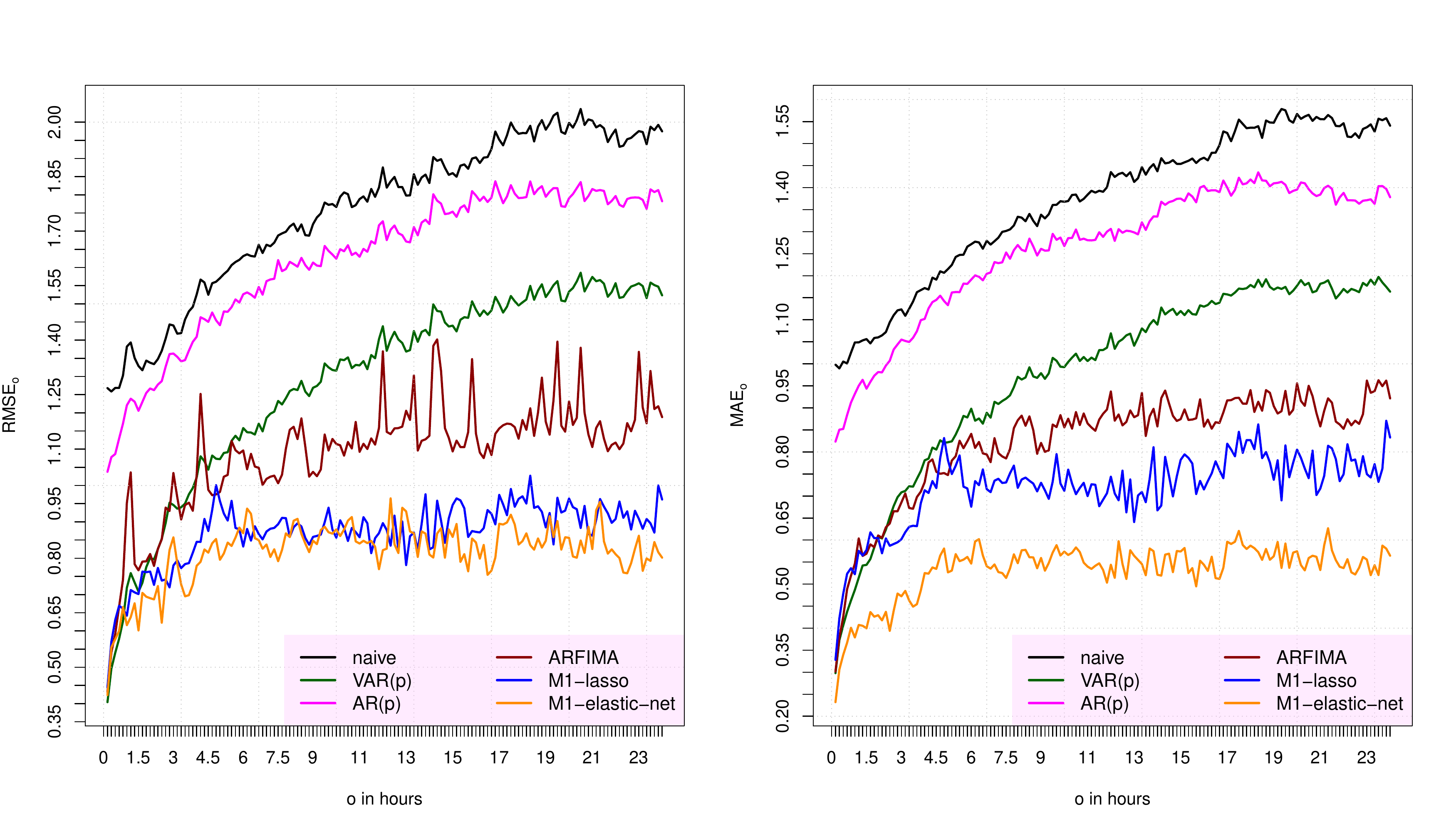}
  \caption{RMSE (first column) and MAE (second column) by forecasting horizon for station Lindenberg for all models.}\label{figure:meas2}
\end{figure}

\begin{figure}[h]
  \includegraphics[width=.5\textwidth,trim= .01cm .3cm .2cm .7cm,clip=true]{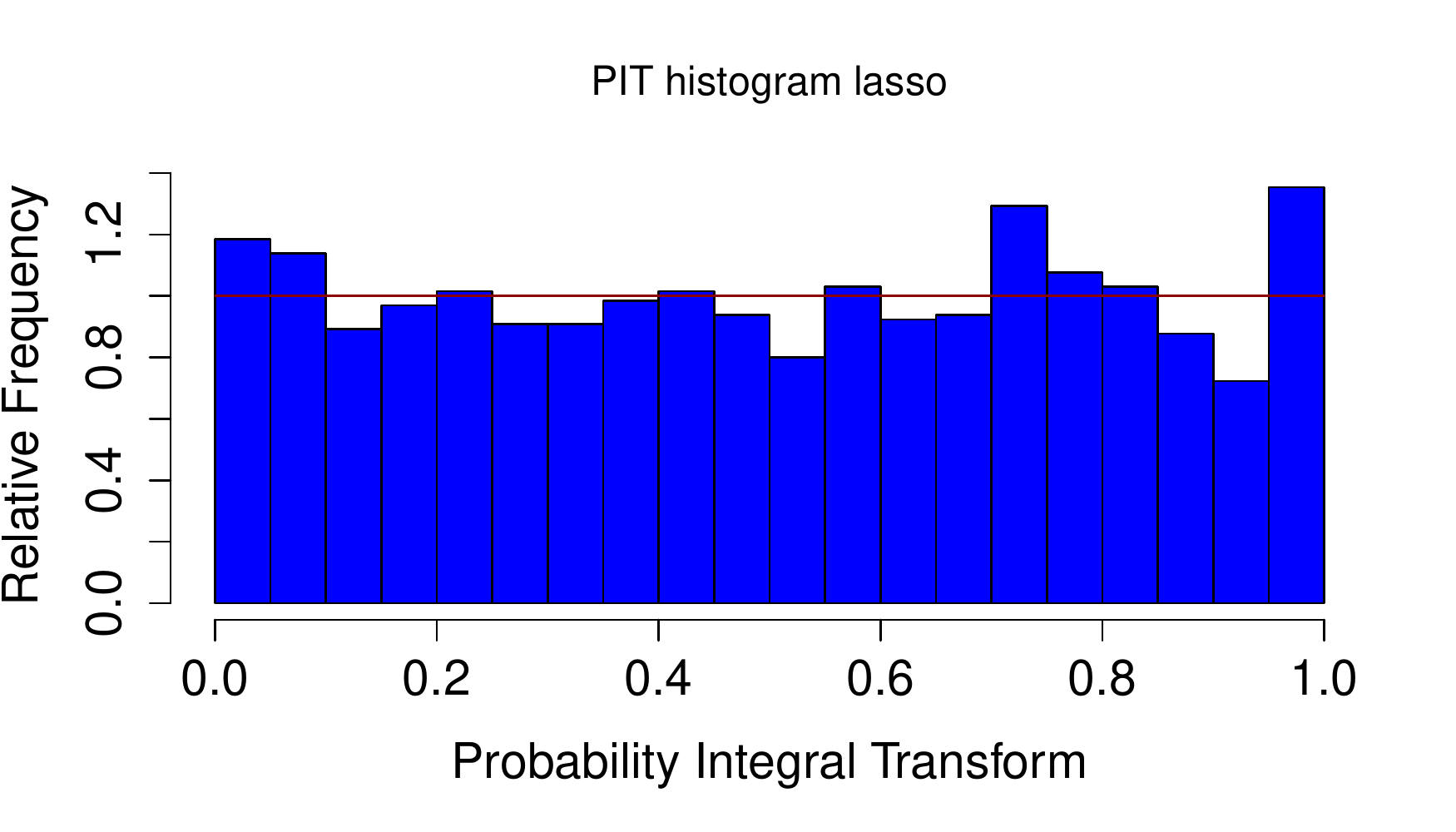}
  \includegraphics[width=.5\textwidth,trim= .01cm .3cm .2cm .7cm,clip=true]{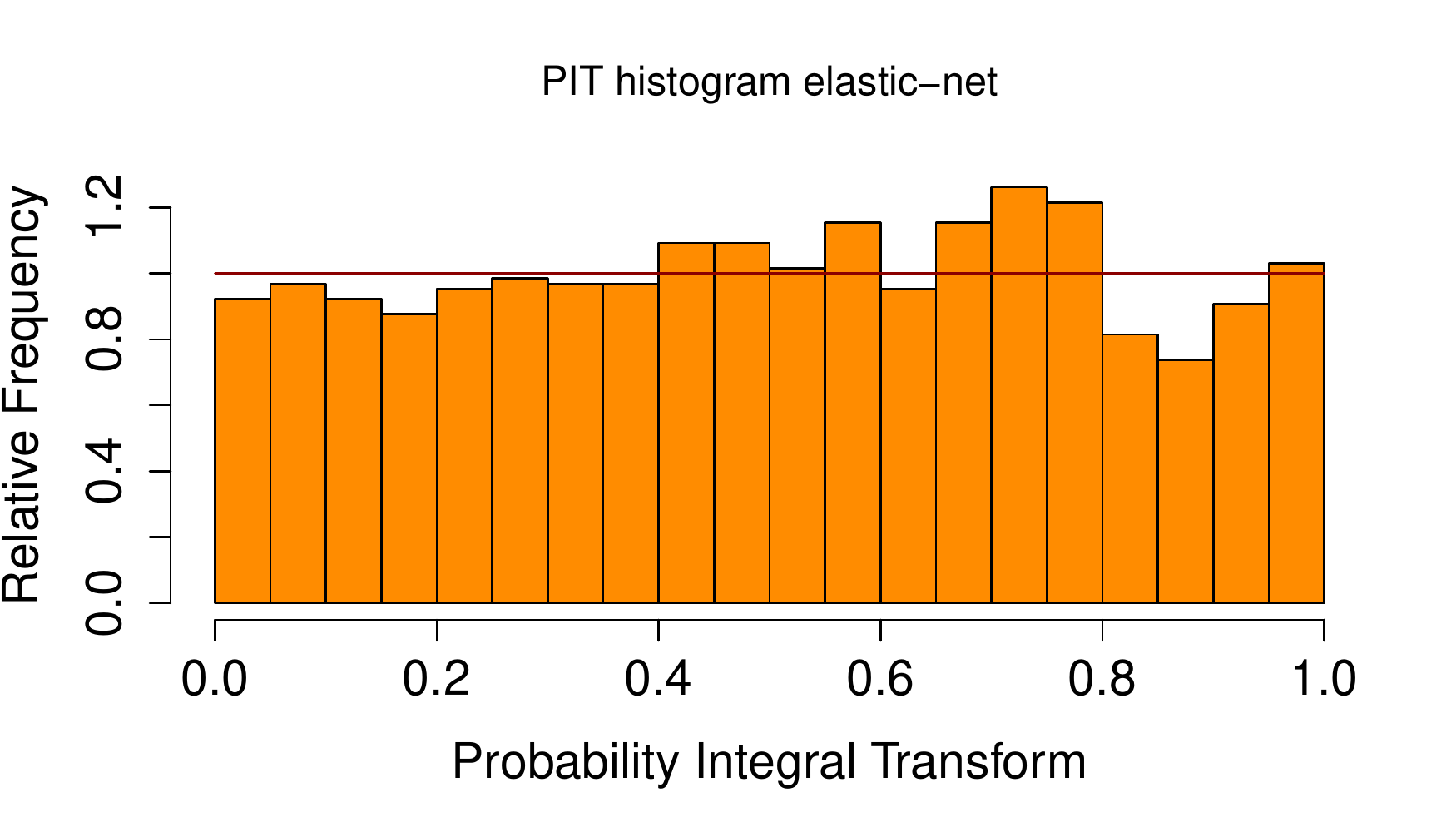} \\
    \includegraphics[width=.5\textwidth,trim= .01cm .3cm .2cm .7cm,clip=true]{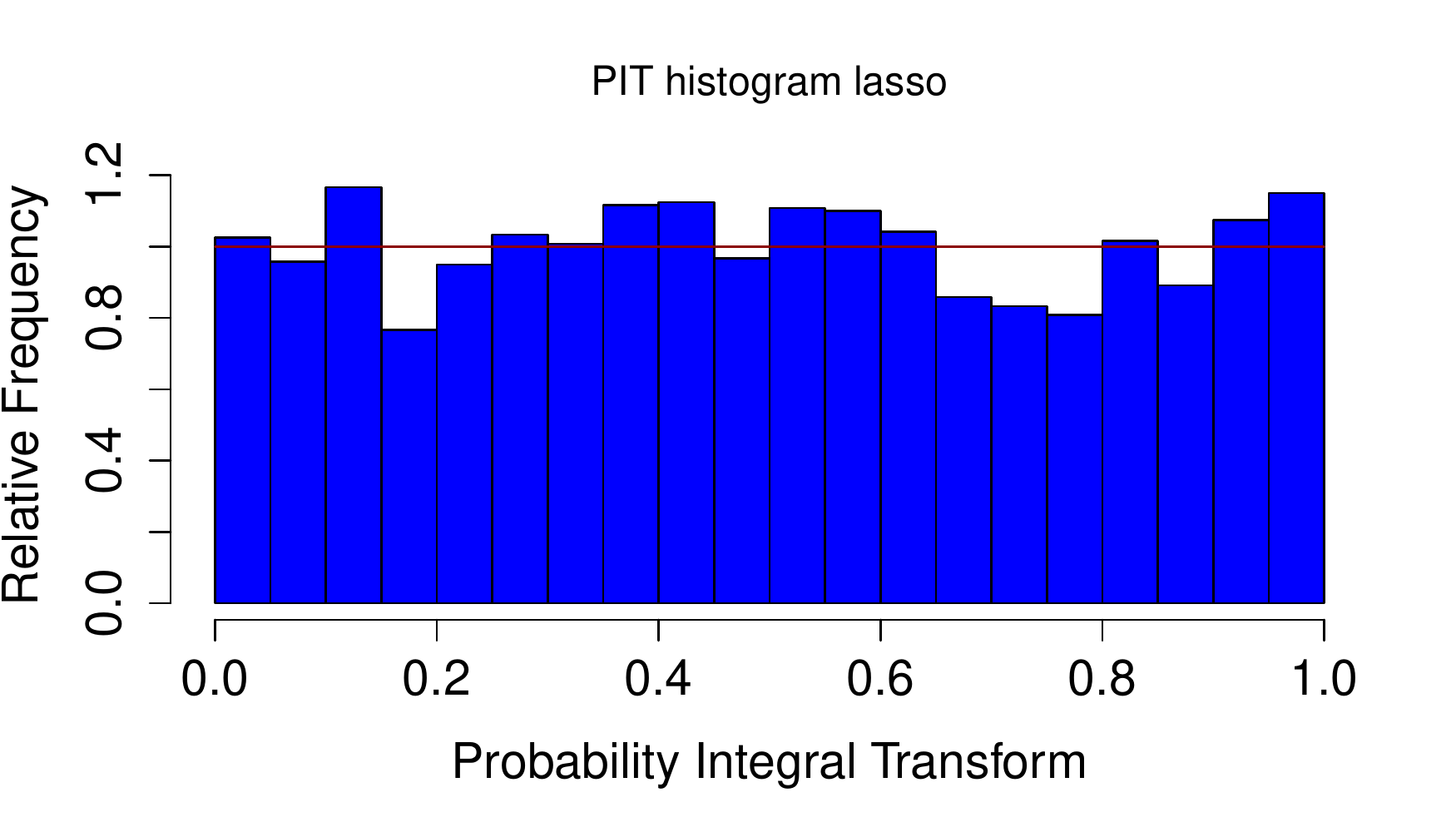}
  \includegraphics[width=.5\textwidth,trim= .01cm .3cm .2cm .7cm,clip=true]{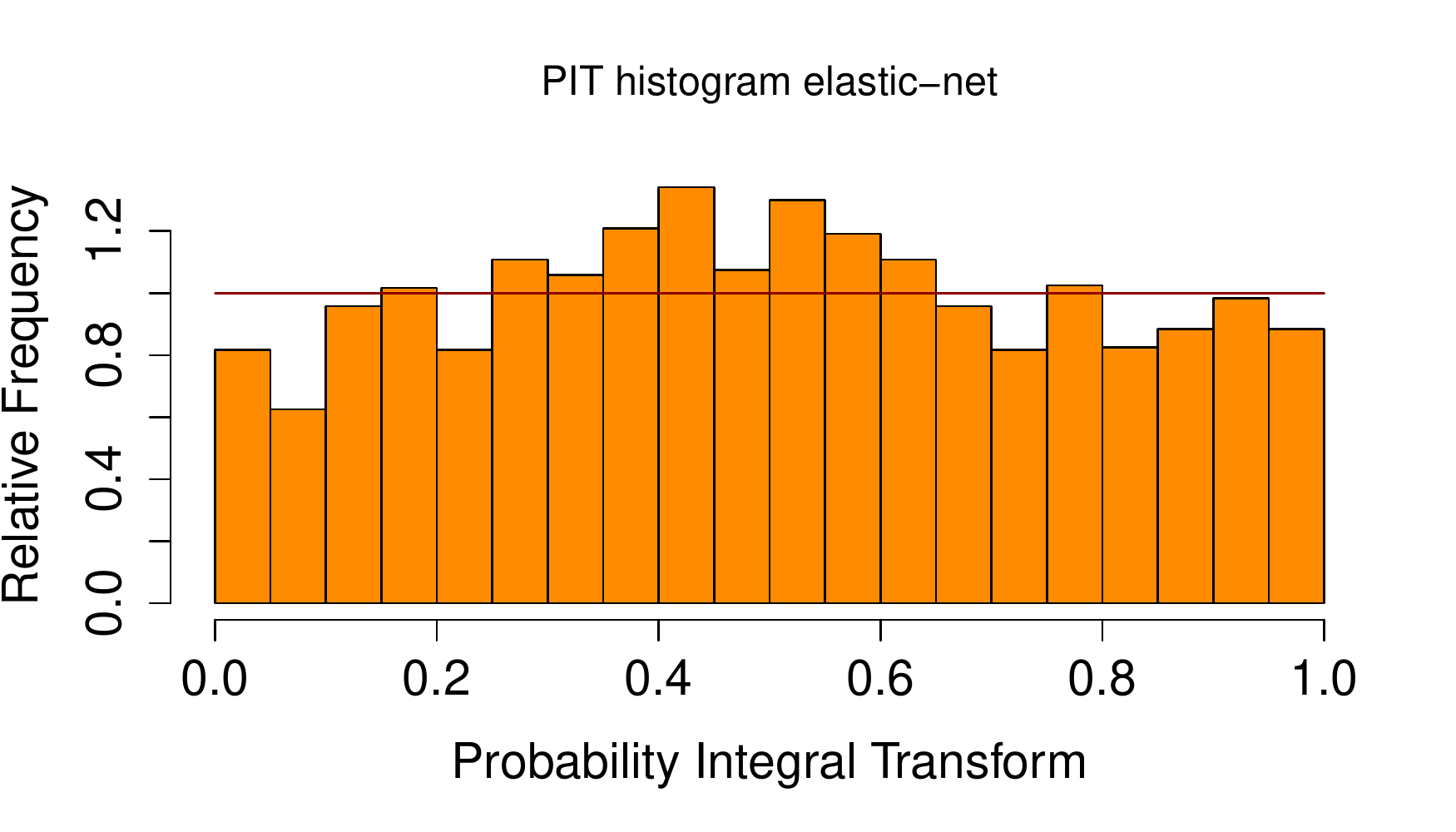}
  \caption{Probability integral transform (PIT) histogram for lasso (first column) and elastic-net (second column) for the forecasting horizon of one hour for station M\"uncheberg (first row) and Lindenberg (second row).}\label{figure:PIT12}
\end{figure}

\begin{figure}[h]
  \includegraphics[width=.5\textwidth,trim= .01cm .3cm .2cm .7cm,clip=true]{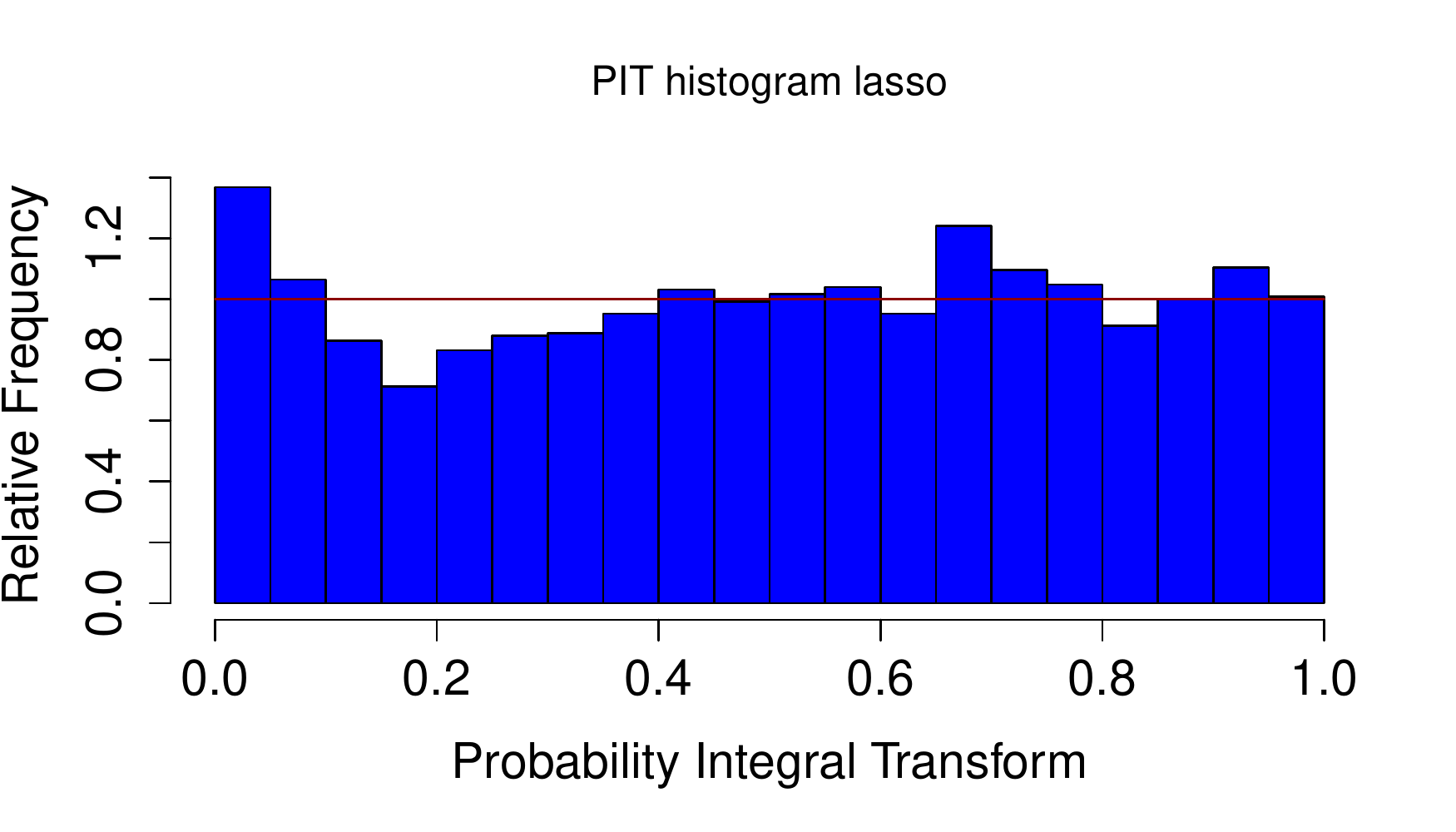}
  \includegraphics[width=.5\textwidth,trim= .01cm .3cm .2cm .7cm,clip=true]{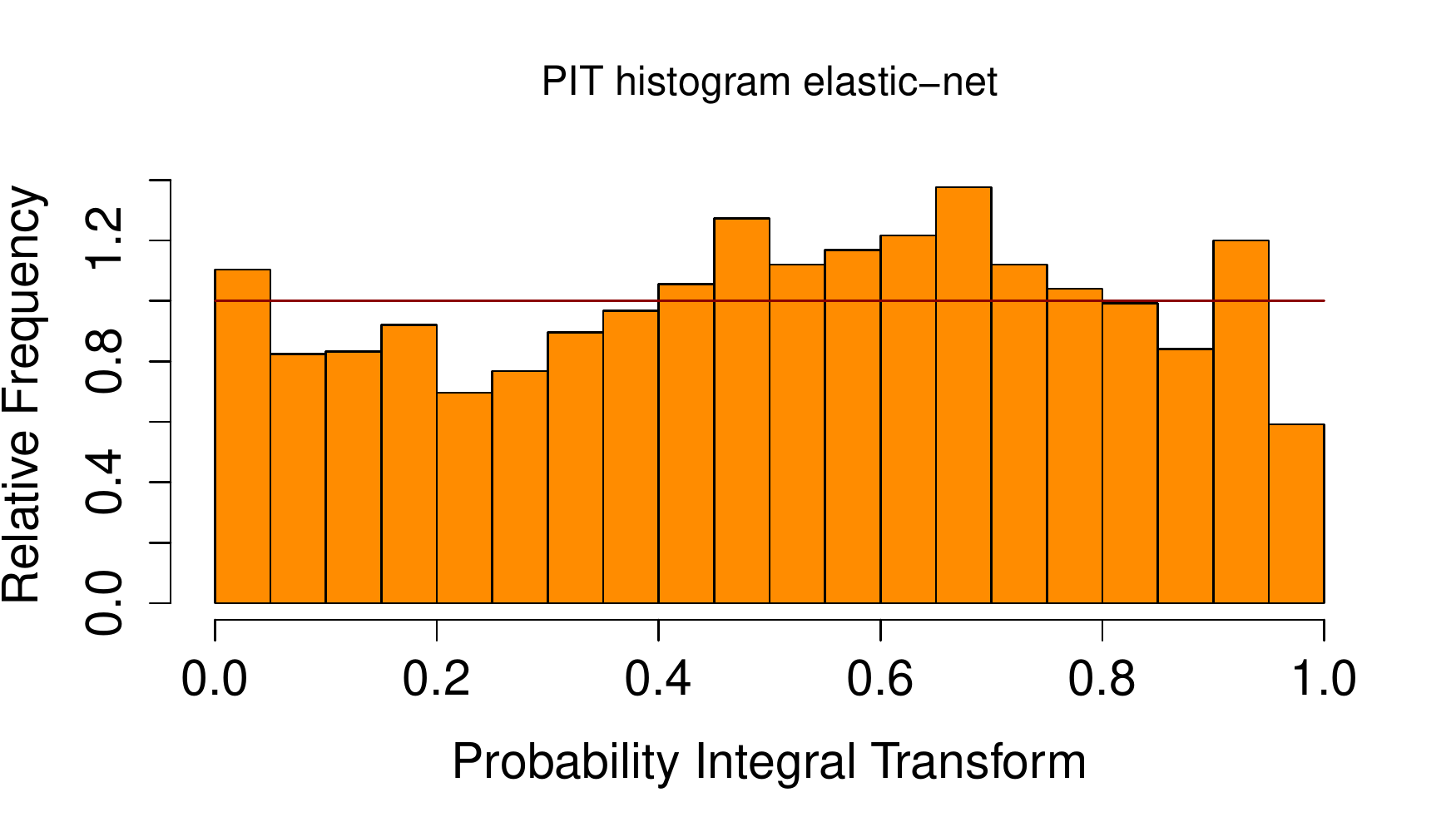} \\
    \includegraphics[width=.5\textwidth,trim= .01cm .3cm .2cm .7cm,clip=true]{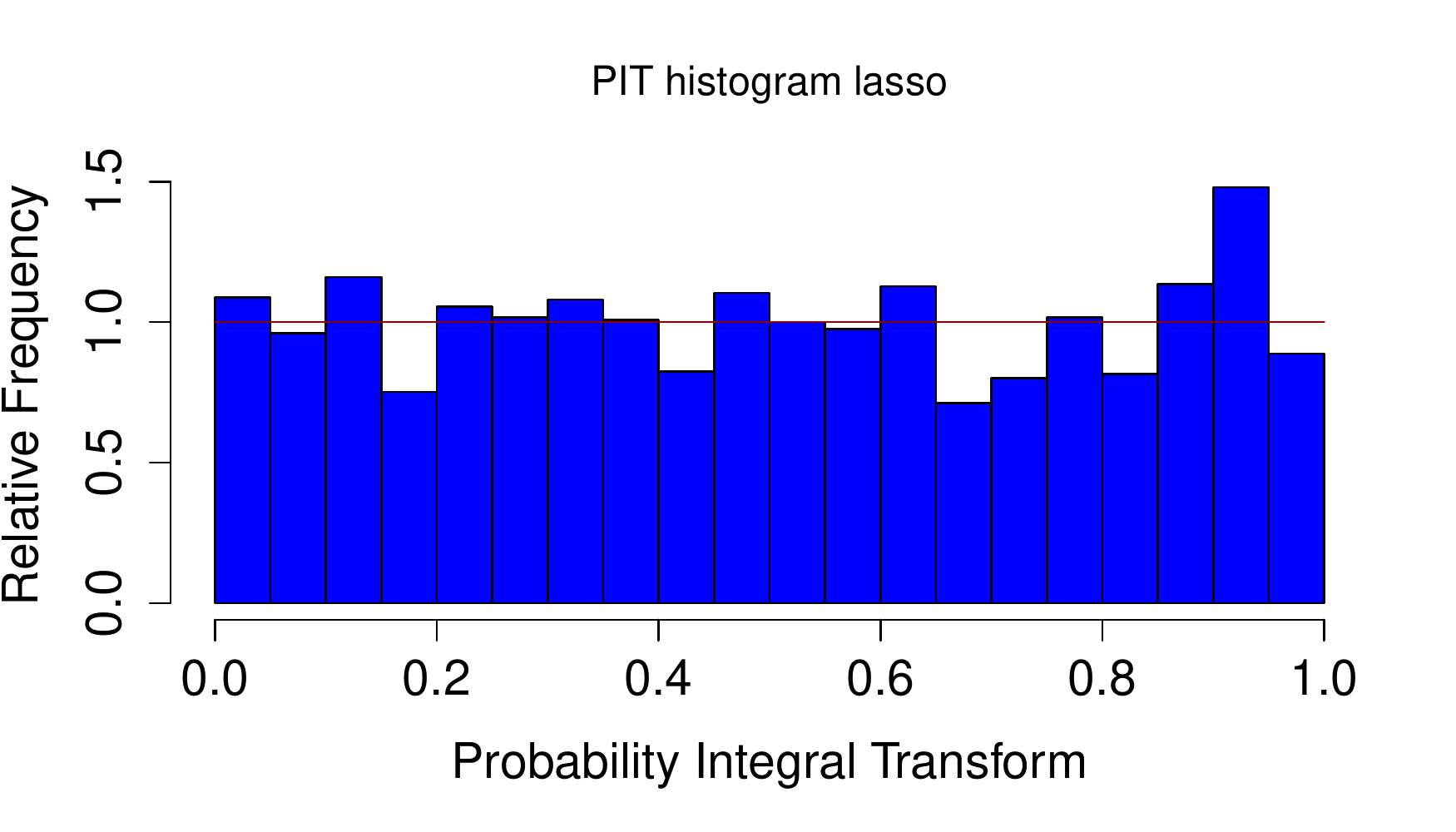}
  \includegraphics[width=.5\textwidth,trim= .01cm .3cm .2cm .7cm,clip=true]{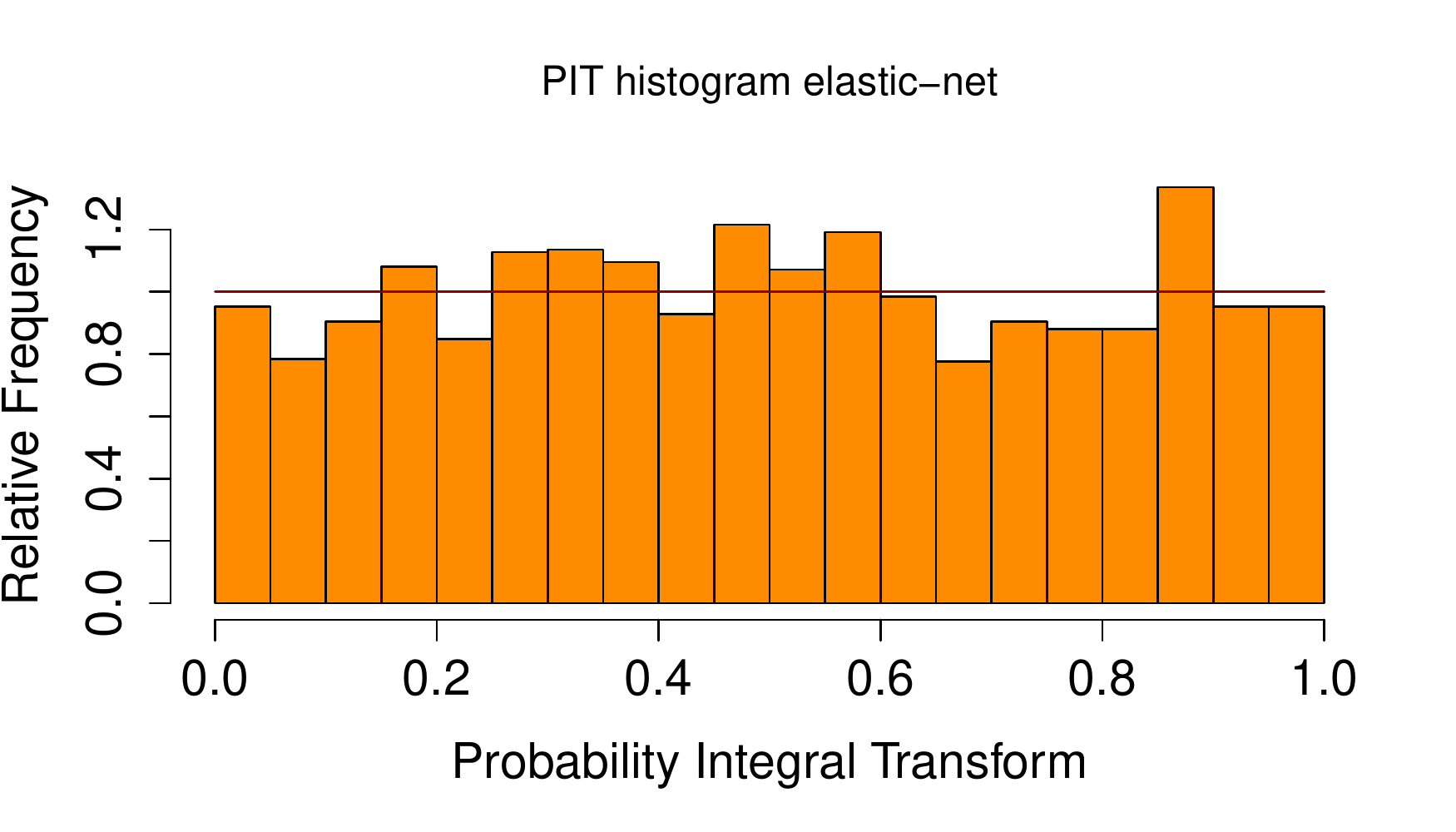}
  \caption{Probability integral transform (PIT) histogram for lasso (first column) and elastic-net (second column) for the forecasting horizon of four hours for station M\"uncheberg (first row) and Lindenberg (second row).}\label{figure:PIT72}
\end{figure}

In this section, we evaluate our model and the benchmark approaches according
to their prediction performance. Common criteria are the root mean
square error (RMSE), the mean absolute error (MAE) and the probability integral transform (PIT) histogram. PIT histograms allow for an insight into calibration and sharpness of forecasts, as \cite{Gneiting2007} point out. The forecasts are
performed for a time frame from July 2011 to December 2011 for
out-of-sample forecasts. We select $N = 1000$ points in time ($\tau^{(i)}, i = 1, \ldots, N$) in the
out-of-sample period at random. Forecasts are calculated at
horizons of up to a maximum of one day (i.e. 24 hours = 144 steps), as 24 hours conclude the short- to medium-term forecasting horizon of wind speed, as, e.g., \cite{Lei2009} and \cite{Giebel2011} point out. RMSE and MAE are calculated by

\begin{eqnarray}
RMSE_o &=& \sqrt{\frac{1}{N} \sum_{i = 1}^N \left(Y_{\mathcal{N},m,\tau^{(i)}+o} - \widehat{Y}_{\mathcal{N},m,\tau^{(i)}+o} \right)^2 } , \\
MAE_o  &=& \frac{1}{N} \sum_{i = 1}^N  \left| Y_{\mathcal{N},m,\tau^{(i)}+o} - \widehat{Y}_{\mathcal{N},m,\tau^{(i)}+o}  \right|   ,
\end{eqnarray}

\noindent where $\widehat{Y}_{\mathcal{N},m,\tau^{(i)}+o}$ is the
$o$-step forecast of wind speed and $Y_{\mathcal{N},m,\tau^{(i)}+o}$ is
the actual observation. Figures \ref{figure:meas1} and \ref{figure:meas2} present the
out-of-sample aggregated forecasting error results. In all cases, our novel model is able to outperform the benchmark models. Mostly, even the
competitive ARFIMA and the extensive VAR(p) model are outperformed
substantially. Additionally, we observe that the na\"ive model
is clearly outperformed in each case. For few shorter
forecasting steps, the highly persistent VAR(p) model returns weakly
lower errors. Indeed, these results are only obtained for the station Lindenberg. As forecasting horizons increase
beyond one hour, our novel model is the overall winner. Finally, we are able to conclude that the iteratively re-weighted lasso and elastic net approach outperform the benchmark methods. Looking at MAE, the elastic net outperforms lasso. RMSE however leaves no distinct result for lasso and elastic net.\\
\noindent Figures \ref{figure:PIT12} and \ref{figure:PIT72} show the PIT histograms for shorter-term forecasting horizons of one and six hours and the iteratively re-weighted lasso and elastic net forecasting technique. As to be expected, we observe that the calibration and sharpness of our forecasts is stable as long as forecasting horizons remain relatively short. However, we observe declining calibration and sharpness for forecasting horizons beyond a scope of six hours. So for medium-term forecasts, we observe a PIT histogram which shows a certain degree of over-dispersion (omitted here to conserve space, available upon request).

\section{Conclusion}\label{section:conclusion}

This article presents a new model for wind speed forecasting. The introduced periodic SVARX-TARCHX model is used to predict the wind speed, wind direction, air pressure and temperature. The model does so by means of non-trigonometric periodic B-spline functions. Moreover, we exploit the spatial distribution of the measurement stations and take conditional heteroscedasticity into account.\\
As the parameters of the sophisticated ARFIMA-APARCH benchmark are estimated by using numerical (quasi) maximum likelihood estimation, it usually takes more than one hour of computing time to find one out-of-sample forecast. In contrast to that, our SVARX-TARCHX model can be identified by modern shrinkage methods like lasso or elastic net, which takes only about six to eight minutes per forecast. Due to our iterative re-weighting scheme, the algorithm is still able to capture the he\-te\-ro\-sce\-das\-tic nature of the observed data.\\
Overall, our model proposition is able to outperform the naive benchmark as well as the VAR. Even the more advanced ARFIMA-APARCH model is outperformed by a significant degree. Modeling periodicity by B-splines instead of Fourier series brings additional flexibility and numerical performance. Results show that the new model captures periodicity quite well. The new model provides a flexible framework with a lot of adjustment options, which makes it a universal tool for many kinds of settings in wind speed forecasting research.


\end{document}